**C:/comets/107P/Draft 7/120510**

# "The 2009 Apparition of Methuselah Comet 107P/Wilson-Harrington: A Case of Comet Rejuvenation?" *

Ignacio Ferrín
Institute of Physics,
Faculty of Exact and Natural Sciences,
University of Antioquia,
Medellín, Colombia
respuestas2011@gmail.com

Hiromi Hamanowa, Hiroko Hamanowa
Hamanowa Astronomical Observatory

Jesús Hernández, Eloy Sira,
Centro de Investigaciones de Astronomía, CIDA,
Mérida, Venezuela

Albert Sánchez,
Gualba Astronomical Observatory,
Gualba, Barcelona, España

Haibin Zhao,
Purple Mountain Observatory, CAS,
2# West Beijing Road,
Nanjing 210008, P. R. China

Richard Miles,
Golden Hill Observatory,
Stourton Caundle,
Dorset DT10 2JP, United Kingdom

Number of pages: 29

Number of Figures: 18

Number of Tables: 6

**Highligths**

- Comet 107P/WH was active in 1949, 1979, 1992, 2005, and 2009 .

- Its age can be measured. We find T-AGE=4700 comet years, WB-AGE=7800 cy.

- This is a methuselah comet very near to dormancy, being temporarily rejuvenated.

- The diameter Deffe=3.67±0.06 km, and the rotational period, Prot=6.093±0.002 h.

- There are three members in the graveyard of comets, 107P, 133P and *D/1891W1*.

**Abstract**

The 2009 apparition of comet 107P was observed with a variety of instruments from six observatories. The main results of this investigation are: 1) 107P/Wilson-Harrington was found to be active not only in 1949 but also in 1979, 1992, 2005 and 2009. The activity is very weak and produces only a slight brightness increase above the nucleus. ($A_{SEC}$ = $V_{NUC}(1,1,0)$ - m(1,1) < 1 magnitude). Since the amount of solar energy received by the object at perihelion has been monotonically increasing since 1928, we conclude that the comet has been active at all apparitions ever after. The probability that the activity in 1949 or in 2009 was due to a surface impact is very small. 2) The rotational period has been determined. We find $P_{ROT}$ = 6.093±0.002 h. The amplitude of the rotational light curve (peak to valley) is $A_{ROT}$ = 0.20±0.01 mag in V. From this data the minimum ratio of semi-axis is $(a/b)_{MIN}$= 1.20±0.02. The rotational light curve is identical to the shape of a saw tooth. The shape of the object must be peculiar and *has sharp edges.* 3) This is the first time that the phase curve of a comet has been determined in three colors, B, V and R. We find B(1,1,α) = 16.88(±0.06) + 0.041(±0.001)α, V(1,1,α) = 16.31(±0.04) + 0.043(±0.001)α, and R(1,1,α) = 15.90(±0.04) + 0.039(±0.001)α. 4) From the phase curves color indices can be deduced when the comet is not active. We find V-B= 0.57±0.07 and V-R= 0.41±0.06. 5) Assuming a geometric albedo $p_V$=0.04 and the absolute magnitude from the phase plot, we find an effective diámeter free from rotational modulation $D_{EFFE}$=3.67±0.06 km. 6) The secular light curves are presented. The comet is slightly active above the nucleus line ($A_{SEC}$(1,LAG) = 0.36±0.10 mag), but did not exhibited a coma. Activity started +26±1 d after perihelion. Using the definition of photometric age, we find T-AGE= 4700 (+6000, -1700) cy (comet years), an exceedingly old object, a methuselah comet. 7) In the 2009 apparition no tail and no coma were detected using more sensitive detectors than on 1949. This is good evidence to conclude that this object is very near extinction and belongs to the *graveyard of comets.* In a Remaining Revolutions vs Water-Budget Age plot, we have identified the region of the graveyard, and 107P is a member of this group. 8) Using the amplitude of the secular light curve ($A_{SEC}$) vs Diameter ($D_{EFFE}$) diagram (Figure 15), we show that *107P is the most evolved object in the current data base of comets.* In the classification scheme of Ferrín (PSS, 58, 365-391, 2010), *107P is a methuselah comet (WB-AGE = 7800 > 100 comet years), medium size (1.4 < $D_{EFFE}$ = 3.67 < 6 km) nucleus, fast rotating ($P_{ROT}$ = 6.093 h < 7 h), belonging to the* graveyard of comets *(the region with $A_{SEC}$ < 1 mag in Figure 15 or 1.000 cy < WB-AGE in*

*Figure 18 ), that must be very near its dormant phase, were not for the fact that this comet is temporarily being rejuvenated due to a trend of apparitions with decreasing perihelion distances.* 9) Since the general flow of the sample is down and to the left in the $A_{SEC}$ vs $D_{EFFE}$ diagram, and since this flow has been going on for centuries, and since dwarf comets evolve much more rapidly than large comets, *the existence of a significant population of dormant and extinct comets located in the lower left region of the diagram is predicted. This is the graveyard. Three comets have been identified as members of the graveyard, 107P, 133P and D/1891W1 Blanpain.*

## 1. Introduction

In support of the Marco Polo Mission (Review Board, 2010), we have carried out observations of comet 107P/Wilson-Harrington, and compiled its secular light curve (SLC) during the 2009 apparition.

Three previous papers have found no activity in this object (Trigo-Rodríguez et al., 2010; Khayat et al., 2010; Ishiguro et al., 2011). In the present investigation we will present evidence to conclude that the comet was active in the 2009 apparition, and in apparitions 1979, 1992, and 2005 though very weakly (amplitude of the secular light curve above the bare nucleus, $A_{SEC}$ < 1 magnitude), where

$$A_{SEC} = V_{NUC}(1,1,0) - m(1,1) \tag{1}$$

where $V_{NUC}(1,1,0)$ is the nuclear magnitude at R=Δ= 1 AU with R the sun-comet distance and Δ the comet-Earth distance, and m(1,1) is the visual magnitude at R=Δ= 1 AU. $A_{SEC}$ is a measure of activity and a proxy for age.

The history of this object is described in Kronk's Cometography (2010). The object was seen active in 1949 in two Palomar Observatory Sky Survey plates, the red and the blue (Fernandez et al., 1997). There is no other observation in the literature registering the comet as active. Osip et al. (1995) determined the rotational state of the object, finding a double peaked light curve with a rotational period of 6.1±0.05 h that we will confirm.

This paper derives the secular light curve of this comet. Secular light curves (SLCs) have previously been published by Ferrín (2005a, 2005b, 2006, 2007, 2008, 2010a, 2010b, from now on Papers I-VII). In the *Atlas of Secular Light Curves of Comets*, (Paper VI), the preliminary SLC of this object was published. The picture that emerges is that we are dealing with a very old comet that exhibits activity only post-perihelion for a short period of time. The activity is very weak and produces only a slight brightness increase above the nucleus. In Paper IV we have defined a photometric age for comets, T-AGE (or time-age because it is deduced from the time plot), that has been scaled to human ages and it is measured in comet years, cy. It turned out that the preliminary photometric age that was determined was T-AGE= 760±40 cy, placing the object into the methuselah class of comets (those that have T-AGE > 100 cy). Thus we must be dealing with an object near extinction. Its physical and photometric properties are thus of great interest since the extinction of comets is a poorly known subject, and objects in this category are rarely available for study.

There is another peculiarity about this comet, and that is that except in 1949 it never exhibited a tail. All other images exhibited a stellar nucleus devoid of coma or tail. The present investigation fills a gap in the photometric portion of our knowledge.

The comet has an orbital period of 4.285 y or 1565 d, and the perihelion distance is ~1.0 AU and variable. This value makes this object a NEO, a Near Earth Object. There are apparitions when the comet approaches Earth to ~0.3 AU. 2009 was one of those occasions. Even so 107P never exceeded 16$^{th}$ magnitude and thus it was not observed visually.

**2. Observing circumstances.**

The comet was recovered remotely from GRAS (Global-Rent-A-Scopes) Mayhill observatory (MPC code H06) by Guido Sostero (2009) on May 19.4, 2009, at magnitude R= 19.2.

It is interesting to compare the present apparition with former ones. This is done in Figure 1 were we present the minimal distance to Earth as a function of time. We see that close apparitions come in two series separated by 30 years. There is the series of 1932, 1962, 1992, and 2022, and there is the series of 1919, 1949, 1979, 2009, and 2039. The 2009 apparition was very favorable. The blue line at the center of the plot gives the value of q, the perihelion distance, amplified by a factor of 10 with respect to 1 AU to see the variations clearly.

The activity of the nucleus is determined by the amount of energy received from the sun. We are interested in knowing if the 2009 apparition was enhanced or depleted. In practice we only need this information in comparison with 1949 when the activity was first detected. The orbital data for all calculations has been taken from the MPC site. In Figure 2 we show the energy enhancement, EE, vs time compared to 1949. The energy received by the comet is proportional to $1/q^2$. We set EE(1949)=100%. The straight line gives the general trend of the EE during this brief period in the life of this comet and we find a temporal enhancement. There have been several apparitions with significant energy enhancements. 1979 had EE = 5%. 2009 had EE = 2.5% and this was enough to create activity but not to create a tail or a coma. Current detection methods are superior to those of 1949 but did not detect large structures. A good apparition to detect activity will be 2018 when the object will receive 7.1% energy enhancement. However the comet will be at 1.7 UA distance from Earth. A very favorable apparition to detect activity will be 2022, when EE = 7.9% and the distance to Earth will be 0.40 AU (vs 0.38 AU in 2009).

**3. Observational Data Sets**

Observations were carried out from six observatories. Observations from the McGraw Hill (MDM) observatory, USA, MPC code 697, using the 1.3 m reflector telescope coupled with a 4K CCD, were carried out by Jesús Hernández. The scale was 0.6276"/pix. He performed time series photometry on four nights, November 20, 21, 22 and 23, 2009. We performed a dual reduction of this data set: an absolute calibration in B, V, R and I, using the calibration zone 3C474.3 from Gonzalez-Perez et al. (2001), and a differential calibration using the USNO SA2.0 catalog. The agreement is excellent and the results appear in Table 1. A stack of 37 images with a total exposure time of 27 minutes is shown in Figure 3. A profile of the comet and a nearby star shows that the comet did not exhibit evidence of coma. In Table 2 the colors derived from this data set are compared with colors from the literature. However using only these high accuraccy observations, it is possible to conclude that the comet was active, as is evident from the phase plots (Figures 11, 12, 13).

Observations from the National Observatory of Venezuela (ONV), MPC code 303, using the 1 m Schmidt telescope situated at 3600 m of altitude were carried out by Nidia Lopez. The telescope was used in the drift scan mode with a clear filter. The response of the CCD has a maximum in the red part of the spectrum best matched by a broad band R filter. The scale was 1.03"/pix. Two nights were allocated with this instrument but only one night was useful. 3 raster scans were made, containing each 4 exposures of 2.4 minutes each for a total exposure time of 28.8 minutes. The results are listed in Table 3 and the images are shown in Figure 4.

Observations from Purple Mountain Observatory, P. R. China, MPC code D29, using a 1.04 m Schmidt telescope coupled with a 4Kx4K CCD, were carried out by Haibin Zhao. This observer took 12 1-minute exposures using a g' filter. The images were stacked and differential photometry was carried out using NOMAD stars. The stacked images on the comet are shown in Figure 5. The profile of the comet and the stars are compared in Figure 6 and we conclude that they are indistinguishable. His photometry is listed in Table 3. Since his measurement was made with a g' filter which is intermediate between the blue and the visual, we have added 0.16 mag to convert to the B band, and subtracted 0.17 mag to convert to the V band.

Observations from Gualba Astronomical Observatory, Spain, MPC code 442, were made by Albert Sánchez using a 35 cm Meade LX-200 Schmidt-Cassegrain instrument, coupled with a ST9XE CCD and R Cousin filter. His observations appear in Table 3. The catalog used was the Carlsberg Meridian Catalog that provides a dense grid of 95 million stars in the r'-band from DEC -30º to +50º (http://www.ast.cam.ac.uk/ioa/research/cmt/cmc14.html).

Observations from Hamanowa Astronomical Observatory, Japan, MPC code D91 Adati, were carried out by Hiromi Hamanowa, and Hiroko Hamanowa, using a Newtonian telescope of diameter 40 cm, at f/4.5 and a SBIG-ST8 CCD camera. The observations consisted of time series differential photometry using several comparison stars during 12 nights over a total time span of 20 days. The observations were carried out at 5600 A, thus corresponding to the V band. This data set was used to determine the rotational period of the comet. The data is compiled in Table 4 and the rotational light curve appears in Figure 7.

Observations from the 2m Faulkes telescope, UK, MPC code E10, were performed by Richard Miles on two dates and in two bands and are listed in Table 3.

**4. Other Data Sets**

Unfortunately the comet never got within range of a visual magnitude. Juan Gonzalez (private communication) searched for the comet with a 20 cm telescope on Nov. 17-18, reaching to magnitude 15.2. He was not able to detect the comet. The observed magnitude at that time was 16.37 by Faustino Garcia (private communication). Thus unfortunately there are no visual observations of this comet in the International Comet Quarterly, ICQ, in the 2009 apparition (Green, 2010).

The MPCOBS site maintained by the Minor Planet Center, MPC, compiles all the astrometry on asteroidal and cometary objects. As a byproduct, the magnitude in the R, V and C (no filter) pass-bands are listed. This is a valuable resource. However the problem with this data set is that astrometric photometry is of low quality. Thus this data set is only useful *to complement other photometry*. It is of interest to analyze this data set more fully to see if variability of the comet can be detected. We will adopt a typical error of ±0.2 mag for this data set.

The cometas_obs data set, is owned by a group of Spanish amateur astronomers. They follow a particular procedure of analyzing the image using several square apertures of 10x10 up to 60x60" of arc, thus providing up to six measurements per image. The total magnitude of the comet can then be found by extrapolating to an infinite aperture (Paper III). In this particular case the object did not exhibit a coma, thus the 10x10 arc second aperture is the appropriated one to use. Additionally they use either the USNO SA2.0 catalog or the Calsberg Meridian Catalog, CMC. The 14$^{th}$ edition of this Catalog covers the declination range from -30º to +50º. CMC claims an error of ±0.17 mag at the level of magnitude 17$^{th}$ in the R photometric band, making the photometry very accurate.

**5. A warning on asteroid photometry in the 1970s.**

Carl Hergenrother (private communication) sent a warning on asteroid photometry in the 1970s: *"It was not uncommon for MPC Code 675 (Palomar, Helin) estimates to be up to 2.5 magnitudes too bright. From the NeoDys site it is possible to identify the 3 brightest measurements of 107P = 1979 VA as those made on 1979 November 15.18,*

*16.11 and 16.22 with magnitudes 11.0, 11.0 and 12.0. The last observation is from observatory 688 (Lowell, Giclas). Luckily there were photoelectric observations by Alan Harris done at nearly the same time and published in Harris and Young (1983): 1979 November 16.10, magnitude 13.25. This observation is the mean brightness after accounting for rotational variations. It suggests that the photographic magnitudes are too bright by ~2 magnitudes".*

The same comment seems to be applicable to MPC Code 688 (Lowell, Giclas). Thus the three bright observations mentioned above must be removed from the secular light curve of this object presented in Paper VI.

**6. Rotational Light curve**

The determination of the amplitude of the rotational light curve is an important issue in this work for the following reason. In the *Atlas of Secular Light Curves* (Paper VI) it is shown that this object is a methuselah comet (photometric age, P-AGE > 100 comet years), one that exhibits very weak activity above the nucleus line. Comet 28P/Neujmin 1 exhibited a coma with V(observed) - $V_{NUC}$ = -3.2 magnitudes, where $V_{NUC}$ is the nuclear magnitude. Comet 133P/Elst-Pizarro did not exhibit a coma with V-$V_{NUC}$ = -2.8 magnitudes (Paper IV). Comet 2P/Encke did not exhibit a coma at aphelion with V-$V_{NUC}$ = -2.4 magnitudes (Paper VI). And comet 107P did not exhibit a coma with V-$V_{NUC}$ = -0.36 magnitudes (this paper). Thus there seems to be an intermediate value at which the coma disappears hidden inside the seeing disk. This quantity is defined as the *Threshold Coma Magnitude* and its estimated value based on the examples shown above is TCM = -3.0±0.2. If the total minus nuclear magnitude is contained within TCM, then the comet is active, the coma is contained inside the seeing disk and can not detected. Consider that at 1 AU, a typical seeing disk of 1.5" corresponds to 1090 km, while the effective diameter of this comet is D = 3.46±0.32 km (Licandro et al., 2009). Thus there is plenty of space for an undetected coma.

Thus in the 2009 apparition we expected to find very low activity, no coma and no tail. How then are we going to detect activity? By measuring the brightness above the nucleus using high precision photometry in the phase plot. However, if the observed magnitude is inside the amplitude of the rotational light curve (peak to valley), then the activity will be confused with a rotational variation. To be certain in declaring the existence of activity, the observed magnitude must be above half the rotational amplitude or the rotational light curve. That is why the amplitude of the rotational light curve is important in this work, and why it is needed in the phase plots discussed below.

During the 1979 apparition Harris and Young (1983) determined a rotational light curve with a rotational period of 3.556 h and amplitude of 0.08 magnitudes. We have re-reduced this data using Phase Dispersion Minimization, PDM, and we were able to retrieve this same period. Additionally we included the night of 1979 November 28$^{th}$, by Robert Millis, and the rotational period changed slightly to a value of 3.7 h and an amplitude of 0.07 magnitudes. The rotational modulation was clearly seen. However these are likely incorrect periods. Reasons for incorrect periods might be under sampling of the data or sampling at a specific rate which obscures the true period, or very low level activity of the nucleus.

Only one comet exhibits a shorter rotational period than 107P, 133P/Elst-Pizarro, with $P_{ROT}$= 3.47 h. All other comets exhibit rotational periods larger than 4.9 h (the record holder, 126P/IRAS, Groussin et al., 2004). Osip et al. (1995) observed 107P object on 1992 December 1$^{st}$ and 2$^{nd}$, and they found a rotational period of 6.1±0.05 h and rotational amplitude of 0.20 magnitudes.

Our own determination using the data from HAO Observatory listed in Section 2 (see Table 4), is presented in Figure 7. We find the rotational sidereal period, $P_{SID}$ = 6.093±0.002 h, and a rotational amplitude peak to valley $A_{ROT}$ (PTV) = 0.20±0.01 mag, with $N_{OBS}$ = 345. This value is in excellent agreement with that of Osip et al. (1995) and thus the rotational periods of Harris and Young (1983) must be spurious. We used Alan Harris' Fortran 77 FALC Fourier Analysis software to calculate this period.

It is worth mentioning that the rotational light curve of 107P is very peculiar and in fact it is indistinguishable from the shape of a saw tooth. Thus the shape of the object must be uncommon and *it has sharp edges.* With a rotational period of only 6 hours, 107P belongs to the class of *fast rotators*. It is also noticiable how much this rotational light curve resembles that of comet 133P.

Recently Urakawa et al. (2011) reported a rotational period of 7.15 hours for 107P. They also found evidence of a second periodicity of 2.38 hours suggesting non-principal axis rotation. They found our period and the Osip et al. (1995) period to be pseudo-periods. The main reason we do not agree with their result is that their light curve has 6 peaks and 6 valleys, when any not spherical rotating asteroid must necessarily produce a basic two peak two valley light curve (like ours does). We recall here Occam's razor: A simple two peak two valley solution (with strength of 102) must be preferred to a complex 6 peak 6 valley solution (with strength of 59). A possible reason why Urakawa et al. had difficulties in finding our period is that they observed from Δt=+44 to +59 d which lies inside the active period (+26 d to +60 d ), while we observed between Δt= +63 to +77 d when the comet was inactive. In othe words, activity erased our period. Our data is very clean and shows only classical random noise.

**7. Colors**

Fernandez et al. (1997) conclude that the activity of the comet in 1949 was due only to gas emission since it was detected mostly in the blue plate of POSS. However West et al. (1979) make this interesting remark: *"….note that the weakness of the tail on the red plate does not necessarily mean that the tail consists of gas only. Even dust tails which shine by reflected sun light, and which are generally redder than gas tails, are normally weaker on red than on blue photographic survey plates because of the different emulsion sensitivity. It is for this reason that comets are much easier to discover on blue-sensitivity plates…This does not apply to comet observations with CCDs, since these detectors are more sensitive in the red spectral region. So we are convinced that minor planet (4015) really had a dust tail in 1949".*

Colors have been compiled from the literature and are listed in Table 2 where we have also included the values determined from this work. It can be shown that the colors determined in this work deviate significantly from former values. We believe this is evidence of activity. This is more clearly seen in Figures 9 and 10 where we plot the color indices B-V vs V-R and R-I vs V-R along with color indices of inactive comets taken from Paper IV. Three comets deviate to the right, 50P, 86P and 152P because they were active at the time of the observation. We find V-B= 0.57±0.07 and V-R= 0.41±0.06.

**8. Phase Plots and Activity**

There are numerous photometric observations of this comet in the literature and they have been compiled into Table 3 alongside with our own observations. From this information it is possible to create the phase diagram in the B, V and R bands, and to decide if the comet was active. The phase diagram plots the reduced magnitude to $\Delta$ = 1 AU, R = 1 AU, versus the phase angle, $\alpha$.

The phase diagram is of great importance in the study of comets, for several reasons: (1) the phase line is defined by the fainter observations that fit a linear law. (2) The best way to determine the absolute magnitude of a comet is by using the phase diagram. The magnitude extrapolated to phase angle $\alpha = 0^o$ (excluding the opposition surge), is the absolute magnitude $m_N$ (1,1,0) where m is the magnitude in a given band. If we know the geometric albedo in the same band then the effective diameter of the comet can be determined. (3) If the phase diagram in several photometric bands is available, then the difference in absolute magnitudes gives the color indices V-B, V-R, R-I, devoid of coma contamination. The difference in two bands at a single epoch do not give the color indice because the slopes of the phase plot in the two bands are different. (4) The observations should lie inside the rotational amplitude of the nucleus. Twice the standard deviation of the linear fit, gives an idea of the rotational amplitude if the photometric errors are small. (5) Values well above the phase line and beyond half the rotational amplitude must be coma contaminated. Thus this is an excellent way to determine activity, and more sensitive than profile fitting by a factor of 10. However this implies that knowledge of the rotational amplitude is needed.

Because of the existence of the Threshold Coma Magnitude, TCM, with a current estimated valued of TCM= -3.0±0.2 mag, a comet may exhibit no coma and still be active. The coma is contained inside the seeing disk. High precision photometry is capable of detecting ±0.1 magnitude enhancements. And *the phase plot is the right place to look for* *these magnitude enhancements.* Thus the phase plot is of great significance in the study of comets.

In Figure 11 we present the phase plot of 107P in the B band. The data comes from Tables 1 and 2. We see that our photometry serves to define the phase plot, since the observations are the faintest that satisfy a linear law. The linear law is very well defined. It is possible to conclude that the comet was active in this band in four days of observation. We find B(1,1,α) = (16.63±0.05) + (0.055±0.004) α. The correlation coefficient CC = 0.993, the standard deviation is σ = 0.07, and the number of observations, $N_{OBS}$ = 3. Observations from this work made in 2009 between Δt = +13 and +44 d show the comet as active. The observation at Δt = +44d is by Haibin Zhao from Purple Mountain Observatory. And the observation by Buie and Picken in 1992, at Δt = -13 d also shows the comet as active on that date.

In Figure 12 we present the phase plot of 107P in the V band. The data comes from Tables 1 and 2. The phase line is well defined and our photometry between Δt = +29 to +32 d is at least 3σ above the nuclear line, and beyond half the rotational amplitude. The conclusion is that the comet was active in the V band in the four observing dates. We find V(1,1,α) = (16.31±0.04) + (0.043±0.001) α. The correlation coefficient is CC = 0.99, the standard deviation is σ= 0.10 and the number of observations is $N_{OBS}$ = 15.

In Figure 13 we present the phase plot of 107P in the R band. The data comes from Tables 1 and 2. The phase line is well defined and our photometry is several σ above the nuclear line, and beyond half the rotational amplitude. The conclusion is that the comet was active in the five observing dates. Additionally data points by Bowell in 1949 at Δt = + 42d and by Tsumura in 2005 at Δt = +21 d show the comet as active. We find R(1,1,α) = (15.90±0.04) + (0.039±0.001) α. The correlation coefficient is CC = 0.99, the standard deviation is σ= 0.12 and the number of observations is $N_{OBS}$ = 18.

Khayat et al. (2010) observed comet 107P looking for activity. They found none on October $22^{nd}$, 2009, the perihelion date, 26 days before the onset of activity. However they were able to determine a phase coefficient of 0.0406±0.0001 mag/deg, but they did not specify in what pass band this value is valid. Their result is consistent with our values 0.043±0.001 in the visual or 0.039±0.001 in the red pass bands.

Trigo-Rodríguez et al. (2010) also observed the comet. In November 28.76 they could not find any activiy on the comet. Photometry was not given but their plot of FWHM shows the comet with a value well above the mean for the stars. So their result is inconclusive and they show one single date.

Ishiguro et al. (2011) conclude that the comet was dormant or inactive in 2009. However the spectra they show taken 55 d after perihelion (the end of the active period) exhibits a slight enhancement at the location of three $C_2$ emission bands and is not flat. These authors have a time coverage with only 6 data points in their Figure 4 (vs 228 in our work). Two of their four data points in the active period exhibit slight enhancements. Also they fitted a 5 parameter Hapke model to the data, which erases the signature of any activity. Any distribution can be flattened by a 5 parameter model.

It is a common procedure to try to derive new insight about the regolith of a comet or asteroid using Hapke's parametrization. However recent results by Shepard and Helfenstein (2007) cast into doubt the whole procedure. These authors could not find evidence that a given set of parameters could describe uniquely a given surface. They did this using photometric data reduced in a blind test. They included sofisticated reduction procedures including modeling of the three term Henyey-Greenstein phase function. The authors conclude that the fault lies with the scattering model using by Hapke which is inadecuate for physical modeling. In conclusion, any reduction based on Hapke's parameter must, at this time, be looked with caution, and this is the reason why we did not attempt it.

We believe the following are good reasons to conclude that the comet was active in 2009: a) our high precision photometry in the R band shows the comet active above the phase line in Figure 13 by -0.51 mag. b) Two independent data sets show an increase in magnitude for 26 d < t, in Figure 14. c) The absolute magnitude in the R band was determined with great precision using 18 data points, R(1,1,0)= 15.90±0.04. The majority of data points in Figure 14 lie above this value. The average magnitude of observations between +26 and +60 d is <R(observed)> = 15.71±0.02. This value is separated from the nucleus magnitude by 5-10 $\sigma$. d) Urakawa et al. (2011) had great difficulty in retrieving our period and this could be explained by the fact that they observed from Δt = +44 d to + 59 d inside the active period (+26 d to +60 d), while our observations were made from Δt = +63 d to 77 d while the inactive period.

**9. Secular light curve**

Using all the available information, it is possible to create the secular light curve of this comet. See paper VI for the SLCs of 27 comets. Since the behavior of this object departs very slightly from that of an asteroid, it is better to plot the reduced magnitude R(1,1,0) vs time with respect to perihelion, Δt. This is shown in Figure 14 for the R band.

In Figure 14 we plot two active dates in 1949 and 2005 taken from Table 2. The comet turns on +26±1 d after perihelion (November $19^{th}$ ± 1d, 2009). The turn off point is rather uncertain. The maximum amplitude of the activity was -0.50±0.15 magnitudes above the nucleus. Using this information and the definition of photometric age (Paper II):

T-AGE = 90240 / [ Asec * Tactive ] (2)

we deduce a T-AGE = 4700 (+6000, -1700) cy, an exceedingly old object, a methuselah comet.

In Figure 15 the same information is presented showing R(1,1,0) vs Log R. Negative logs to the left do not mean values less than 1, but observations before perihelion. This Figure lists the values of the turn on point, $R_{ON}$ = 1.037±0.005 AU, the turn off point, $R_{OFF}$ = 1.71±0.2 AU and the absolute magnitude at the LAG time, m(1,LAG) = 15.5±0.1.

## 10. Effective diameter an geometric albedo

Assuming a geometric albedo $p_V$=0.04 and the absolute magnitude in the V-band from the phase plot, V=16.31±0.04 we can derive an effective diameter free from rotational modulation, $D_{EFFE}$=3.76±0.06 km.

The argument can be inverted to derive an improve geometric albedo. Licandro et al. (2009) made observations of 107P with the Spitzer infrared telescope, and deduced a diameter of 3.46±0.32 km, and an albedo 0.059±0.011 for this object. They used an absolute magnitude $H_V$ = 15.99±0.10 from MPC17270 which departs significantly from our values. We used a least squares fit to 18 photometric observations from the phase plot (Figure 12) to find V(1,1,0)= 16.31±0.04. This value is 0.31 magnitudes fainter than the value used by Licandro et al. and produces a significant change in the value of the albedo. The new revised geometric albedo is $p_V$ = 0.044±0.002 which is more precise and also more exact than the former value.

This revised geometric albedo is in good agreement with the previous result by Campins et al. (1995), $p_J$ = 0.05±0.01 obtained using the Isothermal Latitude Model (ILM) preferred by those authors. However this value was measured in the J band and has to be converted to the V band. From a sample of comets with albedo measured in the J and V bands, we obtain a conversion factor p(V) = p(J) -0.014 ± 0.010. Thus Campins' et al. results is $p_V$ =0.036±0.010, in agreement with our revised value. Ishiguro et al. (2011) found $p_R$ =0.055±0.012. Using a conversion factor p(V) = p(R) -0.007 ± 0.010 we find $p_V$ =0.048±0.014. We will adopt a mean value $\langle p_V \rangle$ = 0.044±0.002 for this comet.

## 11. Water Budget

We define the water budget as the total amount of water spent by the comet in a single revolution. Festou (1986) found a correlation between the water production rate $Q_{H2O}$ and the reduced visual magnitude for many comets of the form

$$\text{Log } Q_{H2O} = a - b \, . \, m(1,1) \quad (3)$$

Subsequently many other researchers have found the same correlation (Roettger et al., 1990; Jorda et al., 1992, 2008; de Almeida et al., 1997). In the latest determination by Jorda et al. (2008) the following values for the parameters a and b were found

$$a = 30.675 \pm 0.007, \quad b = -0.2453 \pm 0.0013 \qquad (4)$$

Using this empirical correlation it is possible to convert the reduced magnitudes presented in the time plot (Figure 14) to water production rate, day by day. The sum of these values from $T_{ON}$ to $T_{OFF}$ gives the water budget in kg

$$WB = \sum_{T_{ON}}^{T_{OFF}} Q_{H2O}(t)\, \Delta t \qquad (5)$$

From equation 5 and Table 6 we conclude that 107P produces 2.07E+07 kg of water, or 0.005% the amount produced by comet 1P/Halley.

Let us define the *water budget age,* thus

$$\text{WB-AGE [cy]} = 3.58\ 10^{+11} / WB \qquad (6)$$

The constant is chosen so that comet 28P/Neujmin 1 has a WB-AGE of 100 cy.

In Table 6 we compare the water budgets of several comets to place 107P in perspective. We list WB, $R_{SUM}$, $A_{SEC}$, the photometric age T-AGE(1,1), the photometric age P-AGE(1,q), the new water budget age, WB-AGE, the water budget in units of comet 1P/Halley, the radius removed per apparition, and the remaining revolutions for δ = 0.5, 1, 3. 107P is at the bottom of the list. WB-AGE = 7800 cy for 107P. This makes 107P a methuselah comet ( T-AGE > 100 cy). Compare this value with a WB-AGE = 0.13 cy for comet Hale-Bopp, a baby comet. The WB scaled to comet 1P/Halley is also of great interest. 107P spent 0.004 % of the WB of 1P.

The comets listed in Table 6 span a range of six orders of magnitude in WB-AGE and four orders of magnitude in T-AGE.

**12. Mass loss and remaining time**

From Table 6 it is possible to deduce that the comet lost $2.07 \times 10^7$ kg of water in the 2009 apparition using Jorda's calibration. However we are interested in the total mass loss. To calculate it we need the dust to gas mass ratio, δ. Sykes and Walker (1992) favor a mean value δ = 2.9. Due to our ignorance on the value of this parameter for this comet we will explore 3 possible scenarios with δ = 0,5, 1.0, 3.0. With this information it is possible to calculate the thickness of the layer lost per apparition using the formula

$$\Delta r = ( \delta + 1 ) \; WB / 4 \pi r^2 \rho \qquad (7)$$

where r is the radius and ρ the density. The density is given by ρ= ΔM/ΔV. ΔV, the volume removed, is given by $\Delta V = 4\pi r^2 \Delta r$. And ΔM, the mass removed, is given by $\Delta M_{H2O} + \Delta M_{DUST} = WB ( 1 + \Delta M_{DUST}/WB)$. For the density we are going to take a value of 530 kg/m$^3$ which is the mean of many determinations compiled in Paper III. The resulting values of Δr are compiled in Table 6 for 13 comets at one epoch and two epochs for comet 2P/Encke and are shown in Figure 18.

From Table 6 we see that comet 107P lost 0.20 cm in the 2009 apparition. Since the radius of this comet is only 1650 m, the ratio $r_N / \Delta r_N$ = 7E+05 for δ = 1 (see Figure 28). This calculation implies that the comet has been choked by dust, producing suffocation. This value suggests that the comet has reached the end of its evolutionary path, since it is no longer capable of removing significant amounts of the surface layer.

**13. Active, Dormant, Extinct, and Rejuvenated Comets**

Since it has been shown that 107P is a very old object with very weak activity, it is convenient to clarify into what final state this comet is going.

To be an *extinct comet*, the thermal wave has to penetrate inside the nucleus and sublimate *all the ices* residing therein. Herman and Weissman (1987) have shown that the typical thermal wave gets exhausted at a depth of only ~250 m at 1AU for known comet materials. Thus at 1 AU this is the radius of a comet for which the thermal wave has penetrated down to its core and presumably all volatiles have been exhausted after several perihelion passes. Thus to be extinct, a cometary nucleus has to be very small (~<250 m in radius) so that all volatile substances could have been sublimated. *Thus truly extinct (dead) comets are scarce, small, faint, hard to find and characterize and are mascarading as asteroids.*

On the other hand *dormant* comets should be very abundant. When the thermal wave penetrates inside the nucleus, it sublimates all the volatile substances within a few meters depth. After several apparitions there are no more volatiles to sublimate and the comet becomes dormant. *Dormant comets are abundant, large, bright, easy to find and characterize and they are also masquerading as asteroids.*

If the perihelion distance increases, the thermal wave would penetrate less, would not reach to the layer of ices, and the comet would become more dormant. However, if the perihelion distance were to decrease, the thermal wave would be more intense and would penetrate deeper reaching to the ice layer, thus awakening and *rejuvenating* the comet. A new round of activity would ensue. The perihelion distances of comets change randomly due to planetary perturbations as can be seen in Figure 1.

According to the above scenarios and Figure 2, *comet 107P is going into a temporary enhanced active phase (rejuvenation) because of smaller perihelion distances at recent and future apparitions.*

In Figure 16 we plot the amplitude of the secular light curve, $A_{SEC}$, vs the effective diameter, $D_{EFFE}$, for 27 comets. Recalling that $A_{SEC} = V_N(1,1,0) - m(1,1)$, we find that $A_{SEC}$ is not a function of diameter. Small comets are as active as large ones. Goliath comet Hale Bopp is as active as dwarf comet 46P/Wirtanen. Since comets decrease in activity as they age, and since they lose volatiles, comets move down and to the left in this diagram. In other words, *this is an evolutionary diagram.* This is confirmed by comet 2P/Encke that has changed positions between the 1858 and the 2003 apparitions according to the results presented in Paper VI. The location of 107P is at the bottom of the diagram. *This implies that 107P is an extreme object, the most evolved of the whole sample.* In that plot $A_{SEC}$= 0 implies an extinct or dormant comet and 107P is very near to that state.

**14. The graveyard**

Since the general flow of the sample is down and to the left of the diagram in Figure 16, and since this flow has been going on for centuries, and since dwarf comets evolve much more rapidly than large comets, *the existence of a significant population of dormant or extinct bodies located in the lower left region of the diagram is predicted.* We call this region the *graveyard of comets* and we bound it by $0 < A_{SEC} < 1$ mag. Comet 107P is a member of this group.

The graveyard can also be identified in the remaining revolutions vs WB-AGE diagram, Figure 18. Three comets belong to the graveyard region, 107P, 133P and D/1891W1 Blanpain.

**15. Validation of the concept of Photometric Age**

In Figure 17 the envelopes of several comets are compared taken from Paper VI . We see that older comets are nested inside younger objects. In other words, P-AGE and T-AGE classify comets by shape of the SLC. Additionally Figure 17 shows that as a function of age, $A_{SEC}$ and $R_{SUM}$ = ( $-R_{ON}$ + $R_{OFF}$ ) diminish in value. That is, a comet has to get nearer to the sun to get activated, and it is less and less active as it ages.

Additionally, Figure 17 shows that very old comets like 107P and 133P are the only two members of the class of comets with a turn on *after* perihelion. Normal comets have a turn on *before* and a turn off *after* perihelion.

From Figure 17 the conclusion is that $A_{SEC}$ is a measure of the activity of a comet and that activity diminishes monotonically as a function of age.

By chance all comets that have been visited by spacecraft are young objects: 1P/Halley (7 cy), 9P/Tempel 1 (21 cy), 19P/Borrelly (14 cy), 81P/Wild 2 (13 cy), 103P/Hartley 2 (14 cy) (Papers II to VIII). The surface morphology of 107P (3000 cy) is likely to be different and much more evolved than that of previous comets visited by spacecraft, all of which are much younger (Papers II, VIII).

**16. Future apparitions**

The apparition of 2022 and 2039 are favorable. In the apparitions of 2018 the comet will receive a 8.7% energy enhancement. However the distance to Earth is not as favorable as in the apparition of 2022. It is to be seen if the comet will show enhanced activity due to the significant diminution in the perihelion distance on those dates.

**17. Conclusions**

The 2009 apparition of comet 107P was observed with a variety of instruments from several observatories. The main results of this investigation are:

1) *Activity.* 107P/Wilson-Harrington was found to be active not only in 1949 but also in 1979, 1992, 2005 and 2009 though very weakly (amplitude of the secular light curve above the bare nucleus, $A_{SEC}$ = m(1,1) – Vn(1,1,0) < 1 magnitude). Since the amount of solar energy received by the object at perihelion has been secularly increasing in all apparitions ever since 1928, we conclude that the comet has been active at all apparitions ever after. The possibility that the activity in 1949 or in 2009 was due to a surface impact is ruled out.

2) *Rotational period.* The rotational period has been determined. We find $P_{ROT}$ = 6.093±0.002 h. The amplitude of the rotational light curve (peak to valley) is $A_{ROT}$ = 0.20±0.01 mag in V. From this value the ratio of the semi-axis is a/b= 1.20±0.02. The rotational light curve is very peculiar and in fact is indistinguishable from the shape of a sharp saw tooth. Thus the shape of the object is both peculiar and uncommon.

3) *Phase curves.* The phase curves have been determined in three colors, B, V and R, when the comet was not active. We find $B(1,1,\alpha) = 16.88(\pm0.06) + 0.041(\pm0.001)\alpha$, $V(1,1,\alpha) = 16.31(\pm0.04) + 0.043(\pm0.001)\alpha$, and $R(1,1,\alpha) = 15.90(\pm0.04) + 0.039(\pm0.001)\alpha$.

4) *Color indices.* From the phase curves color indices can be deduced when the comet is not active. We find V-B= 0.57±0.07 and V-R= 0.41±0.06.

5) *Color-color diagrams.* In the 2009 apparition the comet exhibited variable B-V, V-R and R-I indices, indicating activity. This is clearly seen in the color-color diagrams, B-V vs V-R and R-I vs V-R which are presented.

6) *Diameter.* Assuming a geometric albedo $p_V$=0.04 and using the absolute magnitude from the phase plot, V(1,1,0)= 16.31±0.04, we find an effective diameter free from rotational influence $D_{EFFE}$=3.67±0.06 km.

7) *Secular light curves.* The secular light curves are presented. The comet is slightly active above the nucleus line ($A_{SEC}$(1,LAG) = -0.36±0.10 mag), and did not show a coma during the whole apparition. Activity started at +26±1 d after perihelion. The active time was 32±15 days. Using the definition of photometric age, T-AGE= 4700 (+6000, -1700) cy (comet years), an exceedingly old object, a methuselah comet.

8) *Final state of evolution.* Using the amplitude of the secular light curve ($A_{SEC}$) vs Diameter ($D_{EFFE}$) diagram (Figure 16), we show that *107P is the most evolved object in the current data base of comets.* In the classification scheme of Ferrín (PSS, 58, 365-391, 2010), *107P is a methuselah comet (T-AGE = 4700 > 100 comet years), medium size (1.4 < $D_{EFFE}$ = 3.6 < 6 km) nucleus, fast rotating ($P_{ROT}$ = 6.093 h < 7 h), belonging to the* graveyard of comets *(the region with $A_{SEC}$ < 1 mag in Figure 16), that must be very near its dormant phase, were not for the fact that this comet is being rejuvenated due to a temporary trend of apparitions with decreasing perihelion distances with time.*

9) *Old population of comets.* Since the general flow of the data set is down and to the left on the $A_{SEC}$ vs $D_{EFFE}$ diagram (Figure 16), since this flow has been going on for centuries, and since dwarf comets evolve much more rapidly than goliath comets, *the existence of a large population of methuselah comets, faintly active, dormant or extinct, of dwarf or small size and fast rotation, and located in the lower left region of the diagram is likely. We call this region the graveyard of comets. Comet 107P is the first member of this group.*

10) *Target's scientific interest.* By chance all comets that have been visited by spacecraft are young objects: 1P/Halley (7 cy), 9P/Tempel 1 (21 cy), 19P/Borrelly (14 cy), 81P/Wild 2 (13 cy), 103P/Hartley 2 (14 cy) (Papers II to VIII). *We predict that the surface morphology of 107P (with a T-AGE of 4700 cy) will be very different and much more evolved that the surface morphology of those comets.* Thus 107P would be a most interesting candidate for a spacecraft mission.

11) The graveyard can be identified in the remaining revolutions vs WB-AGE diagram, Figure 18. Three comets belong to the graveyard region, 107P, 133P and D/1891W1 Blanpain.

12) *Future apparitions.* The apparitions of 2022 and 2039 are favorable. In the apparition of 2018 the comet will receive a 7.1% energy enhancement. However the distance to Earth is not as favorable as in the apparition of 2022. It is to be seen if the comet will show enhanced activity due to the significant diminution in the perihelion distance on those dates.

13) There is an important moral in this work. We had to use different data sets calibrated inhomogeneously to confirm the activity. In future attempts it would be better to use continuous time series differential photometry calibrated homogenously to try to confirm weak activity of a cometary nucleus or asteroid.

**18. Acknowledgements**

HZ thanks the National Natural Science Foundation of China, Grant Nos. 10503013 and 10933004, and the Minor Planet Foundation of Purple Mountain Observatory. We thank Alan W. Harris of Space Science Institute for providing the Fortran software to calculate rotational periods. IF acknowledges Nidia Lopez for her help with the observations at the telescope. He also acknowledges the TAC of CIDA for assigning time to observe this object. We thank Junichi Watanabe for his help in locating Japanese observations of this comet. We thank Carl Hergenrother for his warning on asteroid photometry in the 1970s. We thank Paul Weissman of JPL and Matthew Knight of Lowell Observatory for reading the manuscript critically and suggesting modifications.

Table 1. High Precision Photometry from Observatory 697 and comparison with other data bases.

| Night | B 3C* | V 3C* | R 3C* | R USNOA2* | R COBS* | RMPCOBS* | I 3C* |
|---|---|---|---|---|---|---|---|
| 091120.04 | 17.65±0.11 | 16.46±0.04 | 16.42±0.08 | 16.58±0.14 | 16.00±0.06 | 16.60±0.15 | 15.75±0.07 |
| 091121.04 | 17.64±0.12 | 16.65±0.04 | 16.35±0.03 | 16.36±0.13 | 16.09±0.18 | 16.57±0.38 | 15.97±0.07 |
| 091122.04 | 17.62±0.11 | 16.67±0.04 | 16.20±0.07 | 16.22±0.11 | 16.15±0.09 | 16.30±0.33 | 16.07±0.08 |
| 091123.04 | 17.56±0.11 | 16.70±0.04 | 16.25±0.07 | 16.32±0.13 | 16.41±0.17 | 16.50±0.11 | 16.06±0.03 |

*These dates correspond to Δt= +29 to +32 days after perihelion.
*B3C, V3C, R3C means that the photometry in B, V, or R is based on the calibration zone 3C 454.3 (absolute photometry).
*RUSNO means that the R photometry is based on USNO A2.0 catalog (relative photometry).
*RCOBS means that the R photometry comes from the Cometas_obs site.
*RMPCOBS means that the R photometry comes from the MPCOBS site.
*<R3C-USNO>= -0.06±0.07; <R3C-COBS>= 0.14±0.25; <R3C-MPCOBS>= -0.15±0.06 where <x> is the mean value of x.

Table 2. Colors

| Authors | U-B | B-V | B-R | V-R | R-I |
|---|---|---|---|---|---|
| Harris and Young (1983) | 0.31±0.02 | 0.66±0.02 | - | - | - |
| Harris and Young (IAUC3426) | 0.30± | 0.72± | - | - | - |
| Meech et al. (2004) | - | - | - | **0.406±0.017**[a] | - |
| Meech et al. (2004) | - | - | - | **0.412±0.021**[a] | - |
| Buie and Picken (IAUC5586) | - | - | **0.70±0.06** | - | - |
| Lowry and Weissman (2003) | | **0.61±0.05** | - | 0.20±0.04 | - |
| Lowry and Weissman (2003) | - | 0.75±0.06 | - | - | - |
| **This Work, From Phase Plots** | **-** | **0.57±0.07**[a] | **0.73±0.10** | **0.40±0.06** | **-** |
| This Work 091120 Obs 697 | - | 1.19±0.05 | 1.23 | 0.04±0.04 | 0.67±0.11 |
| This Work 091121 Obs 697 | - | 0.99±0.05 | 1.29 | 0.30±0.04 | 0.38±0.08 |
| This Work 091122 Obs 697 | - | 0.95±0.05 | 1.42 | 0.47±0.04 | 0.13±0.11 |
| This Work 091123 Obs 697 | - | 0.96±0.05 | 1.41 | 0.45±0.04 | 0.19±0.08 |

a. Our photometry is in good agreement with that of Lowry and Weissman in B-V, with Buie and Picken in B-R, and with Meech et al. in V-R.

Table 3. New (This Work) and Old Photometry of comet 107P = 4015 = 1979 VA.

| YYYYMMDD | $t-T_P$ [d][1] | m(R,Δ,α) | B(1,1,α) | R(1,1,α) | V(1,1,α) | R[AU] | Δ[AU] | α[º] | Observers and References |
|---|---|---|---|---|---|---|---|---|---|
| 20091217 | +56 A | 16.77R | - | 17.18 | - | 1.228 | 0.475 | 49.0 | N. Lopez (This Work) |
| 20091205 | +44 A | 16.96 g | 18.78[2] | - | - | 1.152 | 0.405 | 56.1 | Haibin Zhao (This Work) |
| 20091129 | +38 A | 16.24 R | - | 18.10 | - | 1.113 | 0.381 | 61.0 | Albert Sanchez (This Work) |
| 20091127 | +36 A | 16.02 R | - | 17.94 | - | 1.101 | 0.376 | 62.6 | Albert Sanchez (This Work) |
| 20091125 | +34 A | 15.93 R | - | 17.89 | - | 1.090 | 0.372 | 64.2 | Albert Sanchez (This Work) |
| 20091123 | +32 A | 16.05 R | - | 18.05 | - | 1.080 | 0.369 | 65.8 | Albert Sanchez (This Work) |
| 20091116 | +26 A | 16.39 R | - | 18.45 | - | 1.051 | 0.368 | 70.3 | Albert Sanchez (This Work) |
| 20091113 | +21 I | 16.65 R | - | 18.72 | 19.22 | 1.035 | 0.373 | 72.6 | Richard Miles (This Work) |
| 20091104 | +13 A | 17.58 B | 19.58 | - | - | 1.008 | 0.395 | 76.3 | Richard Miles(This Work) |
| 20091104 | +13 I | 16.92 R | - | 18.92 | 19.42 | 1.008 | 0.395 | 76.3 | Richard Miles (This Work) |
| 20050909 | +60 I | 19.1 C | - | 17.4 | 17.90 | 1.27 | 1.49 | 42.0 | ICQ136, Tsumura |
| 20050801 | +21 A | 18.2 C | - | 17.21 | - | 1.03 | 1.32 | 49.3 | ICQ136, Tsumura |
| 20020120 | +299 I | 20.3 R | - | 16.26 | 16.76 | 3.02 | 2.13 | 9.6 | ICQ123, Oribe |
| 20000503 | -403 I | 20.12 R | 16.74 | 15.93 | 16.13 | 3.17 | 2.17 | 1.72 | Lowry and Weissman(2003) |
| 19971230 | +389 I | 21.88 R | - | 16.39 | 16.89 | 3.47 | 2.96 | 15.1 | Meech et al. (2004) |
| 19970112 | +37 I | 18.6 C | - | 17.73 | 18.23 | 1.114 | 1.156 | 51.3 | ICQ120, Nakamura |
| 19960616 | -173 I | 19.2 C | - | 16.68 | - | 2.19 | 1.26 | 13.9 | ICQ124, Nakamura |
| 19960523 | -197 I | 18.9 C | - | 16.05 | 16.55 | 2.37 | 1.35 | 1.1 | ICQ124, Nakamura |
| 19960513 | -207 I | 19.1 C | - | 16.05 | 16.55 | 2.44 | 1.44 | 5.3 | ICQ124, Nakamura |
| 19921026 | +66 I | 18.2 C | - | 17.78 | 18.28 | 1.31 | 0.80 | 49.3 | ICQ124, Nakamura |
| 19920809 | -13 I | 17.7 R | - | 19.08±0.1 | 19.58±0.1 | 1.012 | 0.523 | 75.2 | Buie and Picken (IAUC5586), B-R= 0.7 |
| 19920809 | -13 I | 18.4 B | 19.78±0.1 | - | - | 1.012 | 0.523 | 75.2 | Buie and Picken (IAUC5586), B-R= 0.7 |
| 19881221 | +227 I | 20.5 C | - | 16.69 | 17.19 | 2.58 | 1.93 | 19.1 | Gibson (IAUC 4737) |
| 19791222 | +77 I | - | 17.92 | - | - | 1.41 | 0.48 | 22.8 | Hartmann et al. (1982) |
| 19791222 | +77 I | 16.37 V | - | 16.85 | 17.28 | 1.40 | 0.47 | 21.4 | Hartmann et al. (1982) |
| 19791130 | +55 I | - | 17.53 | - | - | - | - | 17.4 | Harris and Young (1983) |
| 19791130 | +55 I | 14.31 V | - | 16.44 | 16.87 | 1.23 | 0.25 | 15.0 | Harris and Young (1983) |
| 19791117 | +42 I | 13.14 V | - | 16.55 | 16.98 | 1.14 | 0.15 | 17.7 | Harris and Young (1983) |
| 19791116 | +41 I | 13.17 V | - | 16.59 | 17.02 | 1.13 | 0.15 | 18.5 | Harris and Young (1983) |
| 19791116 | +41 I | 13.25 V | - | 16.67 | 17.10 | 1.13 | 0.15 | 18.5 | Harris (IAUC 3426), ±0.05, Arot~0.05, Prot~4h |
| 19491119 | +42 A | 13.75 R | - | 16.73 | - | 1.15 | 0.22 | 37.8 | Bowell (IAUC 5585) |

[1] $t-T_P$ [d] = time with respect to perihelion, I = Inactive, A = Active. Activity in 2000 is doubful.

[2] B filter used by Zhao is intermediate between B and V. We applied a correction of +0.16 mag to transform to the B band and -0.17 mag to transform to the V band.

*Information derived from this Table: R(1,1,α)=15.90(±0.04)+0.039(±0.001)α, V(1,1,α)=16.30(±0.04)+0.041(±0.001)α, B(1,1,α)=16.63(±0.05)+0.055(±0.004)α, B-V= 0.33±0.06, V - R= 0.40±0.11, V = C + 0.11, R = C – 0.32.

Table 4. Rotational measurements of 107P/WH.
Values are corrected for travel time and phase angle. JD = 2.455.000 + ΔJD

| ΔJD | Δm | ΔJD | Δm | ΔJD | Δm | ΔJD | Δm |
|---|---|---|---|---|---|---|---|
| 192.83755 | -0.491 | 194.9576 | -0.591 | 195.03224 | -0.520 | 200.89772 | -0.585 |
| 192.83955 | -0.467 | 194.95912 | -0.596 | 199.84612 | -0.418 | 200.89958 | -0.542 |
| 192.84106 | -0.464 | 194.96064 | -0.574 | 199.84705 | -0.464 | 200.90146 | -0.552 |
| 192.84259 | -0.464 | 194.96216 | -0.565 | 199.84891 | -0.458 | 200.90332 | -0.542 |
| 192.84411 | -0.483 | 194.96368 | -0.603 | 199.85079 | -0.437 | 200.90518 | -0.576 |
| 192.84563 | -0.485 | 194.96522 | -0.563 | 199.85265 | -0.472 | 200.90711 | -0.558 |
| 192.84716 | -0.445 | 194.96674 | -0.542 | 199.85453 | -0.434 | 200.90902 | -0.540 |
| 192.84869 | -0.494 | 194.96827 | -0.495 | 199.8564 | -0.430 | 200.91089 | -0.565 |
| 192.85021 | -0.486 | 194.96979 | -0.584 | 199.85826 | -0.434 | 200.91275 | -0.539 |
| 192.85172 | -0.456 | 194.9713 | -0.559 | 199.86108 | -0.476 | 200.91462 | -0.524 |
| 192.85324 | -0.435 | 194.97283 | -0.517 | 199.862 | -0.522 | 200.91649 | -0.521 |
| 192.85477 | -0.446 | 194.97435 | -0.616 | 199.86387 | -0.535 | 200.91836 | -0.495 |
| 192.85628 | -0.433 | 194.97586 | -0.534 | 199.86574 | -0.470 | 200.92023 | -0.539 |
| 192.85784 | -0.444 | 194.97739 | -0.562 | 199.86761 | -0.535 | 200.92209 | -0.516 |
| 192.85938 | -0.453 | 194.97891 | -0.524 | 199.86948 | -0.495 | 204.90225 | -0.459 |
| 192.86089 | -0.426 | 194.98046 | -0.517 | 199.90203 | -0.535 | 204.90412 | -0.500 |
| 192.86242 | -0.458 | 194.982 | -0.532 | 199.90391 | -0.580 | 204.90598 | -0.460 |
| 192.86394 | -0.447 | 194.98353 | -0.491 | 199.90579 | -0.551 | 204.90785 | -0.436 |
| 192.86545 | -0.476 | 194.98504 | -0.523 | 199.90765 | -0.556 | 204.90972 | -0.479 |
| 192.86698 | -0.428 | 194.98656 | -0.522 | 199.90951 | -0.523 | 204.91158 | -0.498 |
| 192.8685 | -0.438 | 194.98809 | -0.515 | 200.84341 | -0.488 | 204.91346 | -0.447 |
| 192.87001 | -0.469 | 194.9896 | -0.546 | 200.84531 | -0.498 | 204.91532 | -0.433 |
| 192.88068 | -0.472 | 194.99113 | -0.516 | 200.84714 | -0.498 | 204.91812 | -0.445 |
| 192.88222 | -0.519 | 194.99265 | -0.521 | 200.84995 | -0.480 | 204.91906 | -0.478 |
| 192.88374 | -0.510 | 194.99417 | -0.537 | 200.85088 | -0.478 | 204.92092 | -0.450 |
| 192.88525 | -0.497 | 194.99569 | -0.482 | 200.85278 | -0.457 | 204.92278 | -0.430 |
| 192.88678 | -0.514 | 194.99721 | -0.507 | 200.85466 | -0.461 | 204.92466 | -0.456 |
| 192.8883 | -0.543 | 194.99873 | -0.514 | 200.85653 | -0.456 | 204.92652 | -0.461 |
| 192.89343 | -0.617 | 195.00025 | -0.489 | 200.8584 | -0.449 | 204.9284 | -0.453 |
| 192.89454 | -0.539 | 195.00177 | -0.524 | 200.86027 | -0.454 | 204.93026 | -0.448 |
| 192.89675 | -0.603 | 195.00329 | -0.492 | 200.86214 | -0.453 | 204.93212 | -0.474 |
| 192.89896 | -0.577 | 195.00481 | -0.504 | 200.864 | -0.429 | 204.93399 | -0.477 |
| 192.90118 | -0.665 | 195.00634 | -0.542 | 200.86587 | -0.441 | 204.95267 | -0.570 |
| 192.90343 | -0.582 | 195.00785 | -0.442 | 200.86774 | -0.452 | 204.95453 | -0.550 |
| 192.90567 | -0.577 | 195.00941 | -0.469 | 200.86961 | -0.473 | 205.86499 | -0.609 |
| 192.909 | -0.577 | 195.01096 | -0.490 | 200.87242 | -0.445 | 205.86686 | -0.587 |
| 192.9101 | -0.640 | 195.01247 | -0.451 | 200.87334 | -0.452 | 205.86873 | -0.569 |
| 192.91233 | -0.564 | 195.014 | -0.509 | 200.87522 | -0.528 | 205.8706 | -0.582 |
| 194.94085 | -0.583 | 195.01552 | -0.448 | 200.8781 | -0.477 | 205.87247 | -0.598 |
| 194.94238 | -0.588 | 195.01703 | -0.503 | 200.87903 | -0.553 | 205.87433 | -0.583 |
| 194.9439 | -0.627 | 195.01856 | -0.450 | 200.8809 | -0.528 | 205.8762 | -0.553 |
| 194.94542 | -0.572 | 195.02039 | -0.452 | 200.88277 | -0.552 | 205.87807 | -0.581 |
| 194.94694 | -0.598 | 195.0216 | -0.433 | 200.88557 | -0.532 | 205.87995 | -0.515 |
| 194.94847 | -0.588 | 195.02312 | -0.460 | 200.8865 | -0.492 | 205.88182 | -0.609 |
| 194.94998 | -0.600 | 195.02465 | -0.445 | 200.88837 | -0.531 | 205.88369 | -0.564 |
| 194.95151 | -0.612 | 195.02617 | -0.478 | 200.89024 | -0.469 | 205.88555 | -0.548 |
| 194.95303 | -0.592 | 195.02768 | -0.495 | 200.89211 | -0.547 | 205.91242 | -0.531 |
| 194.95456 | -0.630 | 195.02921 | -0.480 | 200.89398 | -0.564 | 205.91335 | -0.553 |

Table 4. Continued.

| ΔJD | Δm | ΔJD | Δm | ΔJD | Δm | ΔJD | Δm |
|---|---|---|---|---|---|---|---|
| 205.91709 | -0.495 | 205.99944 | -0.513 | 206.86381 | -0.561 | 207.00911 | -0.520 |
| 205.91895 | -0.531 | 206.0013 | -0.510 | 206.86603 | -0.608 | 207.01134 | -0.533 |
| 205.92083 | -0.534 | 206.00317 | -0.538 | 206.87268 | -0.598 | 207.01355 | -0.527 |
| 205.92269 | -0.516 | 206.00504 | -0.531 | 206.89261 | -0.562 | 207.01798 | -0.522 |
| 205.92457 | -0.512 | 206.0069 | -0.480 | 206.89482 | -0.570 | 207.02019 | -0.498 |
| 205.92643 | -0.490 | 206.00878 | -0.550 | 206.9137 | -0.561 | 207.02241 | -0.526 |
| 205.9283 | -0.509 | 206.01064 | -0.501 | 206.91592 | -0.521 | 207.02462 | -0.504 |
| 205.93017 | -0.509 | 206.01344 | -0.525 | 206.91813 | -0.521 | 207.02683 | -0.484 |
| 205.93204 | -0.461 | 206.01624 | -0.534 | 206.92035 | -0.507 | 207.02906 | -0.525 |
| 205.93395 | -0.522 | 206.01811 | -0.508 | 206.92256 | -0.515 | 207.03127 | -0.540 |
| 205.93586 | -0.470 | 206.01997 | -0.516 | 206.92478 | -0.545 | 207.03349 | -0.510 |
| 205.93774 | -0.434 | 206.02185 | -0.483 | 206.927 | -0.514 | 207.03571 | -0.488 |
| 205.9396 | -0.520 | 206.02371 | -0.457 | 206.92922 | -0.536 | 207.03791 | -0.503 |
| 205.94146 | -0.441 | 206.02558 | -0.516 | 206.93814 | -0.477 | 207.04013 | -0.552 |
| 205.94333 | -0.468 | 206.02745 | -0.480 | 206.94147 | -0.501 | 207.04234 | -0.482 |
| 205.9452 | -0.501 | 206.02931 | -0.470 | 206.9448 | -0.471 | 207.04456 | -0.464 |
| 205.94707 | -0.447 | 206.03119 | -0.477 | 206.94922 | -0.454 | 207.04678 | -0.463 |
| 205.94893 | -0.469 | 206.03679 | -0.460 | 206.95144 | -0.498 | 207.04899 | -0.503 |
| 205.9508 | -0.427 | 206.03866 | -0.473 | 206.95365 | -0.468 | 207.05121 | -0.460 |
| 205.95268 | -0.416 | 206.05459 | -0.460 | 206.95808 | -0.472 | 207.05342 | -0.493 |
| 205.95457 | -0.482 | 206.05739 | -0.474 | 206.96031 | -0.449 | 211.8713 | -0.479 |
| 205.95643 | -0.468 | 206.05927 | -0.459 | 206.96252 | -0.426 | 211.87241 | -0.523 |
| 205.95829 | -0.501 | 206.06113 | -0.439 | 206.96473 | -0.441 | 211.87462 | -0.493 |
| 205.96017 | -0.549 | 206.063 | -0.451 | 206.96695 | -0.462 | 211.87683 | -0.471 |
| 205.96203 | -0.540 | 206.06673 | -0.420 | 206.96916 | -0.457 | 211.87905 | -0.465 |
| 205.96391 | -0.560 | 206.07047 | -0.467 | 206.97138 | -0.459 | 211.88126 | -0.502 |
| 205.96577 | -0.557 | 206.07234 | -0.431 | 206.9736 | -0.511 | 211.88348 | -0.466 |
| 205.96763 | -0.590 | 206.07422 | -0.434 | 206.97583 | -0.496 | 211.8857 | -0.481 |
| 205.96951 | -0.589 | 206.07609 | -0.409 | 206.97804 | -0.514 | 211.88735 | -0.466 |
| 205.97137 | -0.579 | 206.07982 | -0.420 | 206.98025 | -0.564 | 211.88929 | -0.464 |
| 205.97323 | -0.534 | 206.08542 | -0.449 | 206.98913 | -0.602 | 211.88929 | -0.464 |
| 205.97511 | -0.580 | 206.08731 | -0.427 | 206.99136 | -0.581 | 211.88929 | -0.464 |
| 205.97697 | -0.590 | 206.08918 | -0.424 | 206.99358 | -0.591 | 211.88929 | -0.464 |
| 205.97884 | -0.554 | 206.85053 | -0.434 | 206.99579 | -0.547 | 211.88929 | -0.464 |
| 205.98071 | -0.538 | 206.85274 | -0.478 | 206.99801 | -0.561 | 211.88929 | -0.464 |
| 205.98258 | -0.523 | 206.85495 | -0.493 | 207.00022 | -0.544 | 207.00911 | -0.520 |
| 205.98445 | -0.517 | 206.85717 | -0.442 | 207.00246 | -0.544 | 207.01134 | -0.533 |
| 205.99567 | -0.551 | 206.85938 | -0.503 | 207.00469 | -0.504 | 207.01355 | -0.527 |

Table 5. Confirmed activity of 107P.

| Date | Δt [d] | Band | Amplitude of Activity [mag] | Observer |
|---|---|---|---|---|
| 1949 | +42 | R | -0.65 | Bowell |
| 1979 | +42 to +56 | B-V=+0.66 vs 0.57 | -0.09 | Harris and Young |
| 1992 | -18, -17 | CN Spectroscopy | >-0.75 | Chamberlin et al. |
| 2005 | +21 | R | -0.65 | Tsumura |
| 2009 | +24 to +60 | R | -0.50 | Several, This Work |

Table 6. Water budget, WB, water budget age, WBAGE, Remaining Revolutions, RR = $r_N$ / $\Delta r_N$.
RR=RR(δ) were δ = M(dust)/M(gas). WB-AGE [cy] = 3.58 E+11/ WB [kg] and
P-AGE [cy] = 1440 / [ $A_{SEC}$ . $R_{SUM}$ ] ; T-AGE = 90240 / [ Asec * Tactive ]. $R_{SUM} = -R_{ON} + R_{OFF}$

| Comet | WB [kg] | $R_{SUM}$ [AU] | $A_{SEC}$ (1,1) | T-AGE (1,q) [cy] | P-AGE (1,q) [cy] | WB-AGE [cy] | WB --------% WB(1P) | $r_N$ [km] | $\Delta r_N$ [m] δ = 1 | RR δ=0.5 | RR δ=1 | RR δ=3 |
|---|---|---|---|---|---|---|---|---|---|---|---|---|
| HB | 2.67 E+12 | 52.1 | 11.5 | 0.75 | 2.3 | 0.13 | 590 | 27 | 1.1 | 32730 | 24550 | 12274 |
| 1P | 4.51 E+11 | 18.7 | 10.8 | 4.2 | 6.6 | 0.79 | 100 | 4.9 | 5.6 | 1158 | 868 | 434 |
| Hya | 2.25 E+11 | 6.7 | 11.6 | 22 | 17 | 1.6 | 50.0 | 2.4 | 12 | 272 | 204 | 102 |
| 109P | 1.29 E+11 | 5.9 | 8.2 | 40 | 28 | 2.8 | 28.6 | 13.5 | 0.21 | 84700 | 63500 | 31750 |
| 65P | 3.06 E+10 | 13.0 | 10.8 | 2.8 | 13 | 12 | 6.8 | 3.7 | 0.67 | 7350 | 5500 | 2750 |
| 81P | 2.09 E+10 | 10.2 | 11.4 | 6 | 17 | 17 | 4.6 | 1.97 | 1.6 | 1600 | 1200 | 610 |
| 103P | 1.88 E+10 | 9.9 | 10.7 | 7 | 14 | 19 | 4.2 | 0.57 | 17.4 | 44 | 32 | 16 |
| 2P2003 | 8.58 E+09 | 3.1 | 4.8 | 103 | 64 | 42 | 1.9 | 2.55 | 0.40 | 8580 | 6436 | 3217 |
| 2P1858 | 1.28 E+10 | 3.9 | 6.4 | 61 | 47 | 27 | 2.8 | 3.20 | 0.75 | 11370 | 8525 | 4262 |
| 9P | 1.27 E+10 | 7.7 | 9.0 | 9 | 26 | 28 | 2.8 | 2.75 | 0.50 | 7300 | 5450 | 2700 |
| 45P | 7.95 E+09 | 3.5 | 7.5 | 64 | 40 | 45 | 1.8 | 0.43 | 12.9 | 44 | 33 | 17 |
| 28P | 3.58 E+09 | 4.5 | 3.2 | 100 | 133 | 100 | 0.8 | 11.5 | 0.008 | 1.9E06 | 1.4E06 | 7.1E05 |
| 26P | 1.64 E+09 | 3.2 | 5.2 | 85 | 89 | 218 | 0.4 | 1.47 | 0.02 | 8600 | 6450 | 3200 |
| 133P | 1.81 E+08 | 0.2 | 1.4 | 280 | ----- | 1978 | 0.32 | 2.3 | 0.010 | 3.0E05 | 2.2E05 | 1.1E05 |
| 107P | 2.03 E+07 | ---- | 0.50 | 4700 | ----- | 7800 | 0.004 | 1.65 | 0.002 | 9E5 | 7E5 | 3E5 |
| D/1819 W1 | 1.40E+07 | 3.3 | 2.2 | 240 | 200 | 25000 | 0.003 | 0.16 | 0.16 | 1300 | 970 | 490 |

.

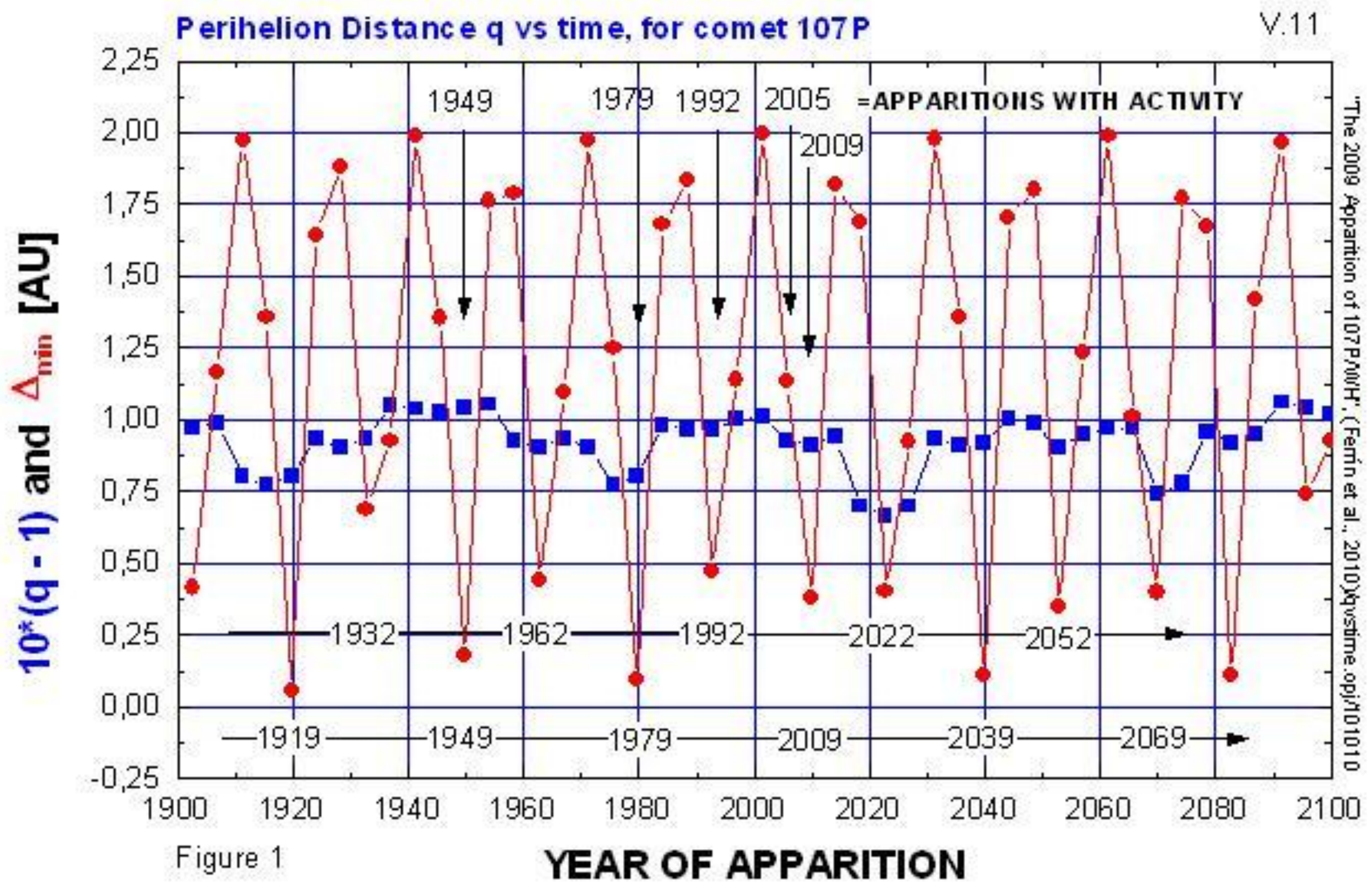

***Figure 1.*** *Minimal distance to Earth as a function of apparition. Close apparitions come in two series separated by 30 years: the series of 1932, 1962, 1992, and 2022, and the series of 1919, 1949, 1979, and 2039. The line at the center of the plot gives the value of q minus 1 AU multiplied by a factor of 10 to see the change in perihelion distance more clearly. There is a trend of diminishing q distances to the sun as a function of time. The arrows indicate the years when the comet was observed to be active (data from Table 5). The data for Figures 1 and 2 has been taken from the ephemeris given at the Minor Planet Center.*

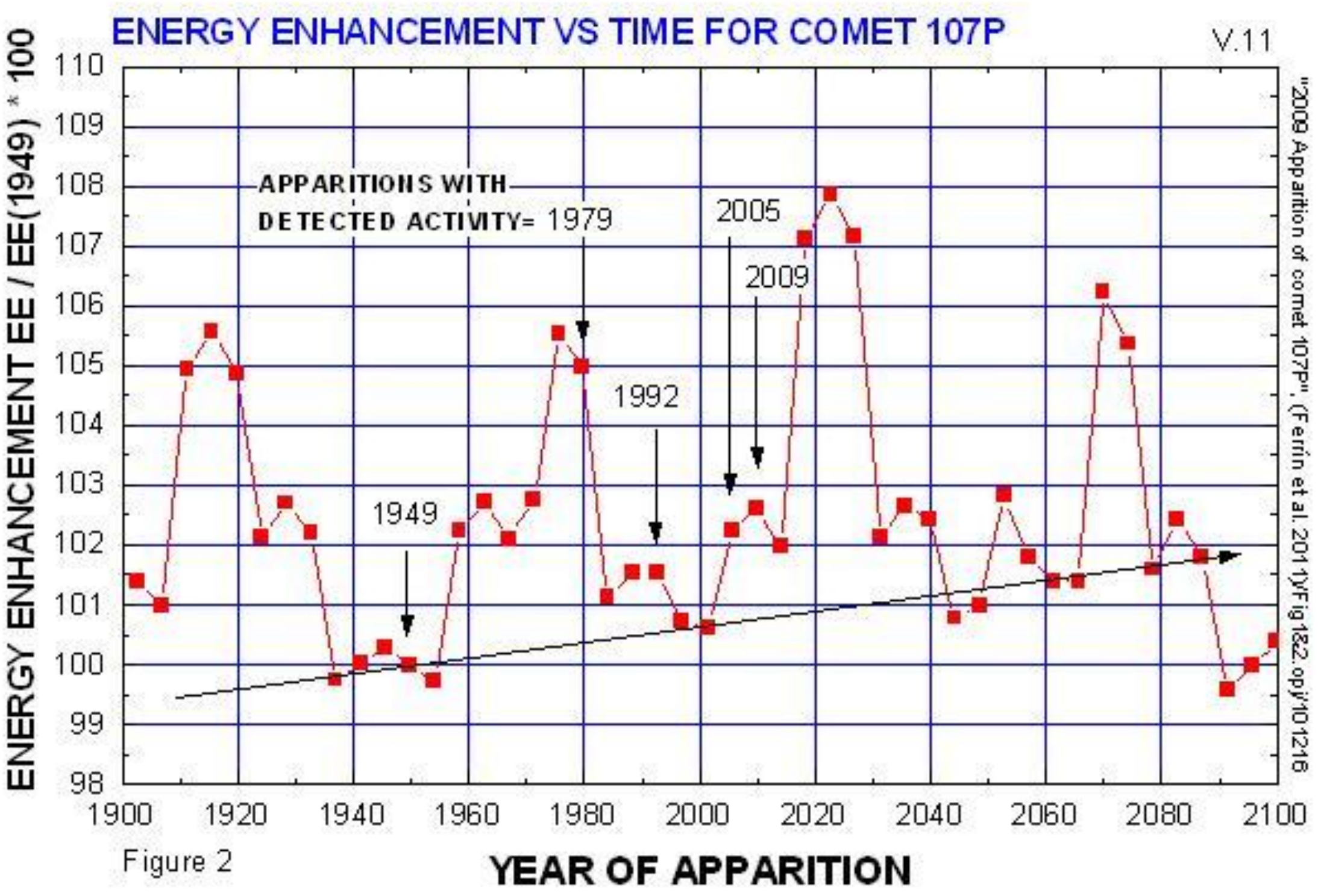

***Figure 2.*** *Energy enhancement, EE, vs time, compared with 1949. The energy received by the comet is proportional to* $1/q^2$. *We set EE(1949)=1. The arrows indicate the years when the comet was observed to be active (from Table 5). Since most years after 1949 show large EEs, this plot makes us to suspect that the comet has been active at all apparitions since 1949. 2009 had EE = 2.5% with respect to 1949, which was enough to create feeble activity. A very favorable apparition to detect activity will be 2022, when EE = 7.9%. The comet is going into a temporary enhanced active phase because all apparitions after 1949 have smaller perihelion distances. This is the evidence to assert that the comet is being rejuvenated.*

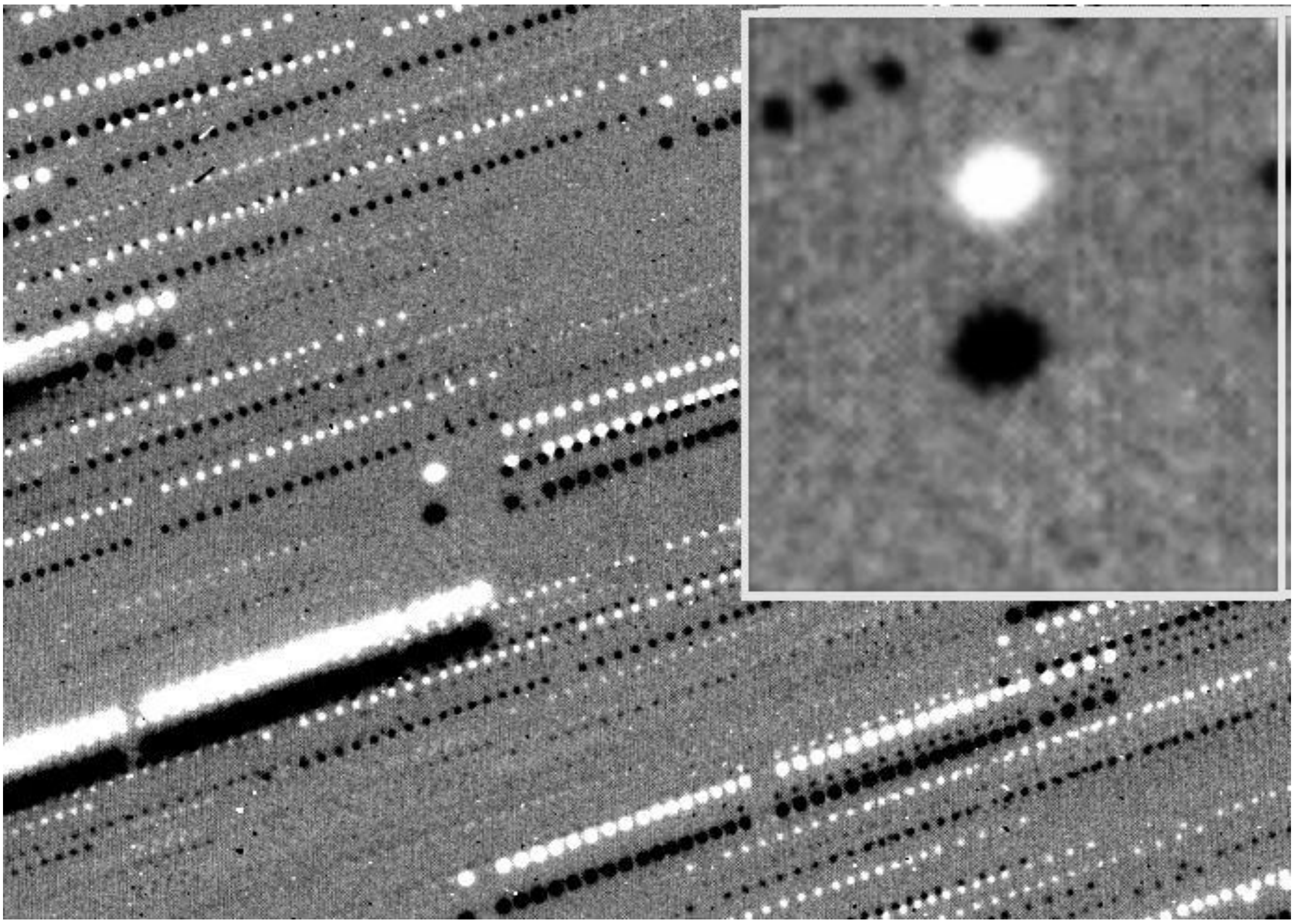

***Figure 3.*** *Composite image of 107P combining all the images from McGraw Hill observatory taken on the night of 091120 ( Δt = +29 d, 3 days after turn on) in the B, V, R and I pass-bands. 37 images combined give a total exposure time of 27 minutes with a telescope of 1.3 m diameter. The enlarged image shown in the inset looks stellar.*

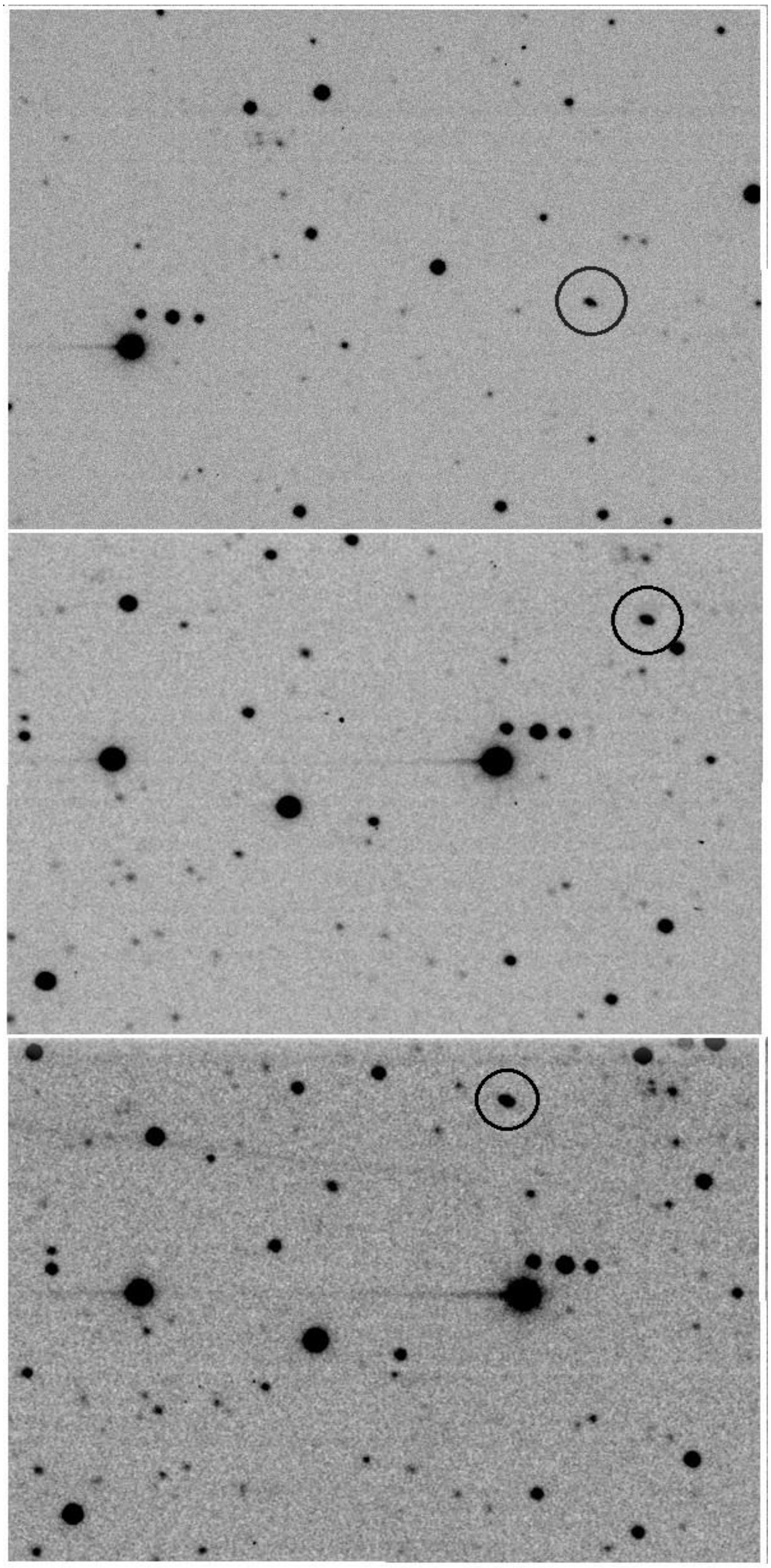

***Figure 4.*** *Three images taken from the National Observatory of Venezuela, with the 1m f/3 Schmidt telescope.  Total exposure time is 7 min*.

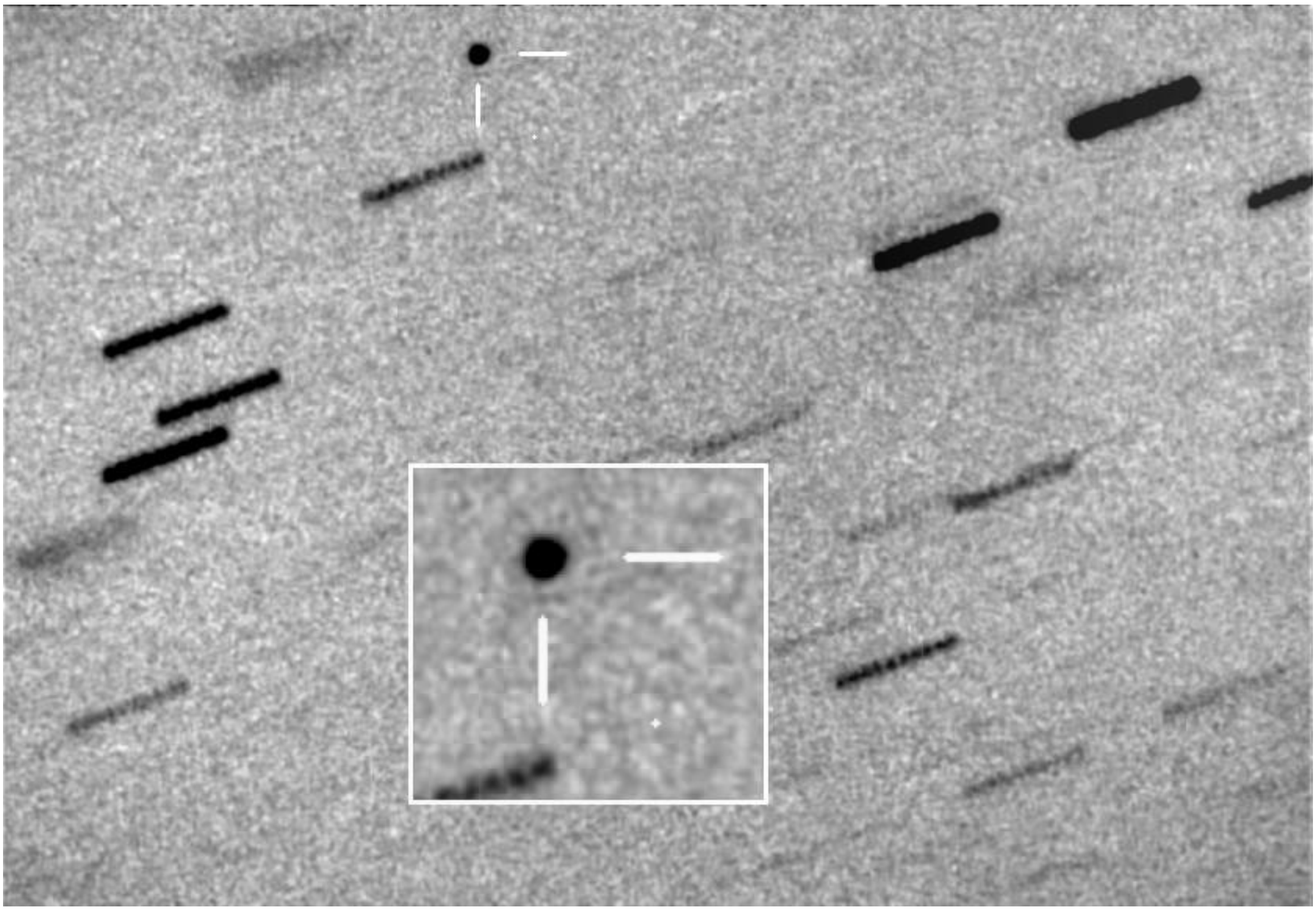

***Figure 5.*** *Composite image of 107P taken from Purple Mountain Observatory. Twelve images taken on 091205 (Δt = +43 d, 17 days after turn on) were stacked for a total exposure of 12 minutes. The comet did not exhibit a coma.*

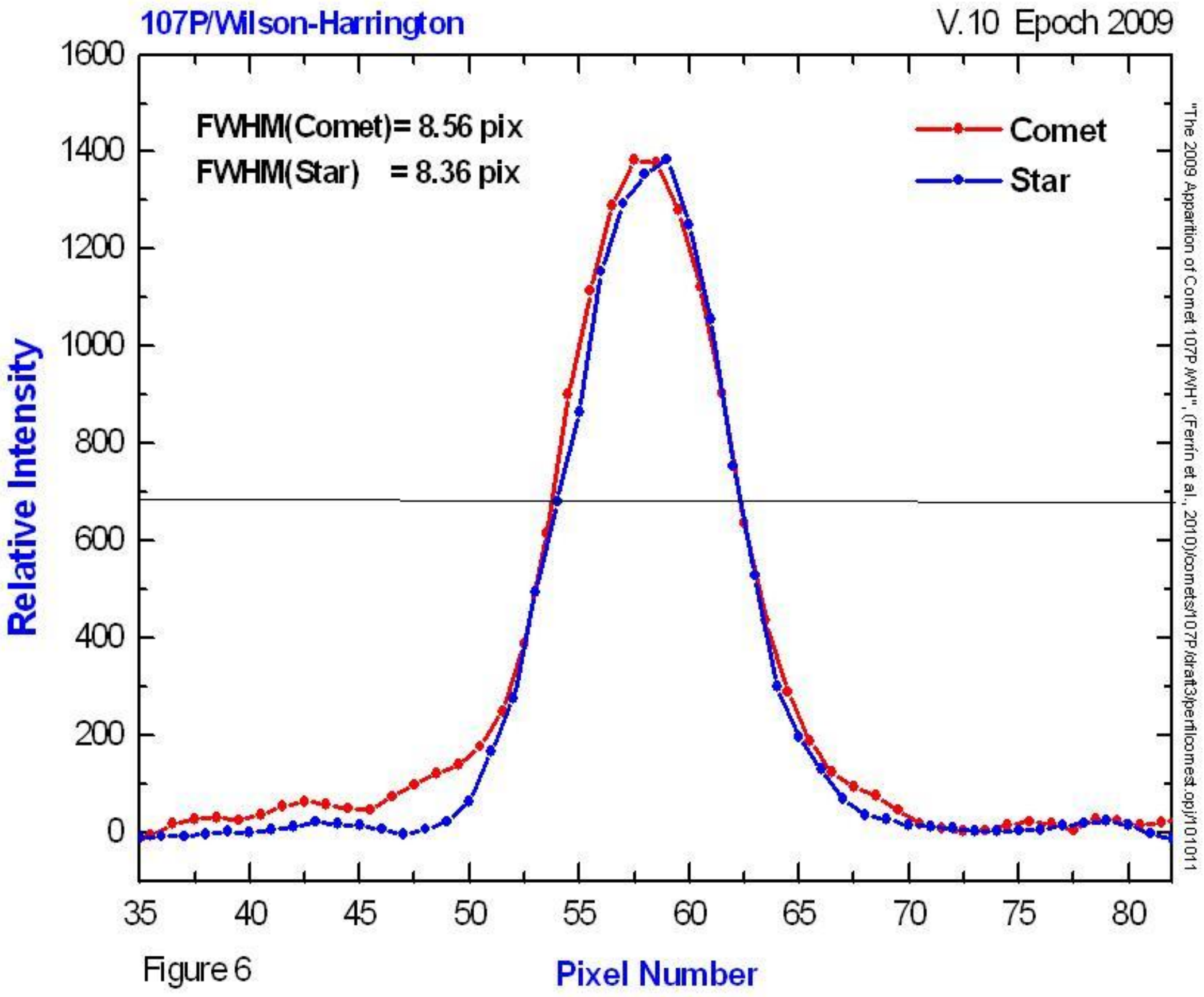

***Figure 6.*** *Profile of the comet and comparison star. The comet is not wider than the star on the night of 091205 (Δt = +43 d, 17 days after turn on). The profile was taken perpendicular to the direction of motion to avoid potential smearing of the image.*

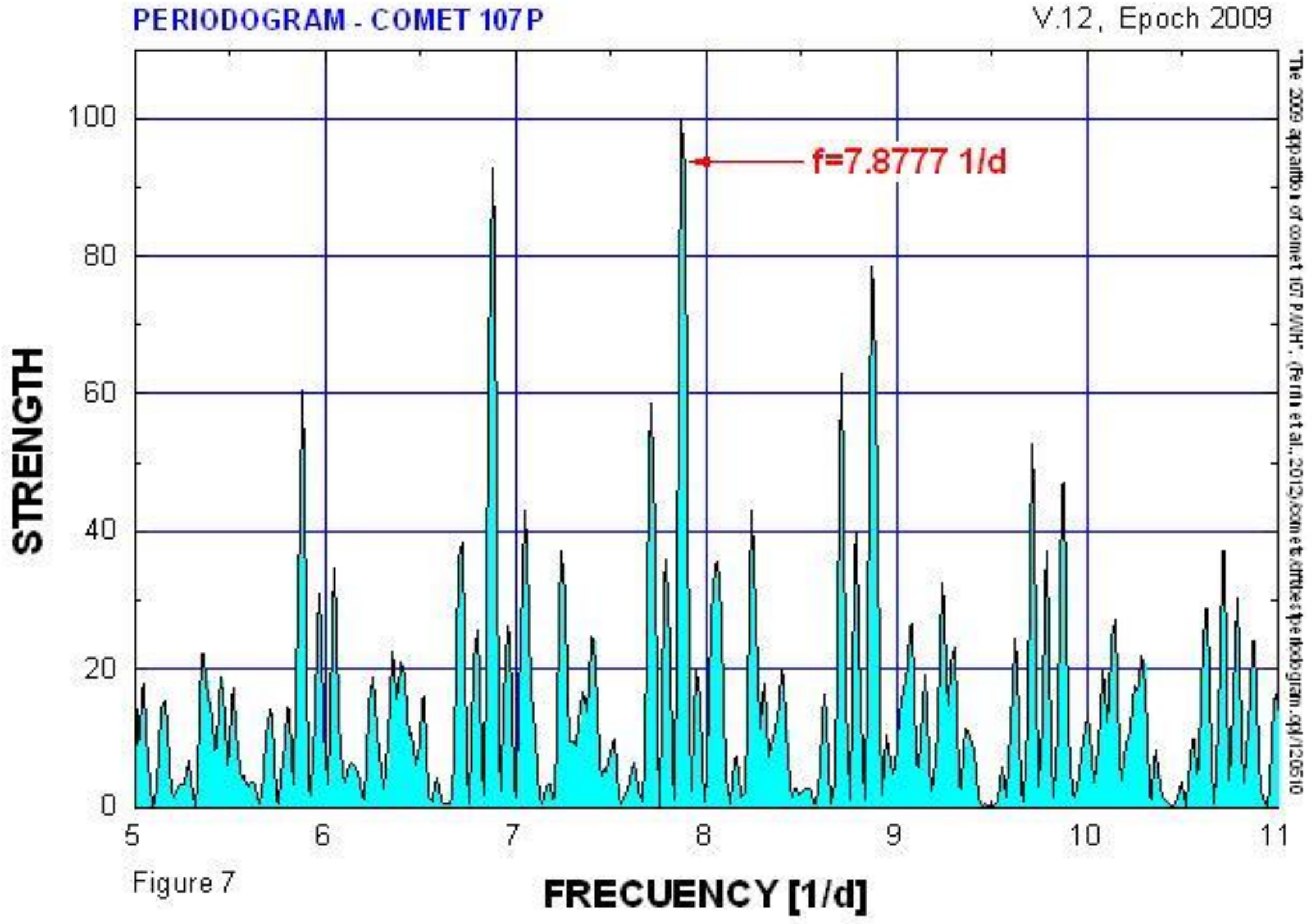

***Figure 7.*** *Periodogram generated with the Discrete Fourier Transform. The most prominent peak at frecuency f=7.8777 1/d with a strength of 102 is the result for a ligth curve with one peak and one valley. The two peak two valley light curve has f/2= 3.93885 which corresponds to a period of 0.253881 d = 6.093 h.*

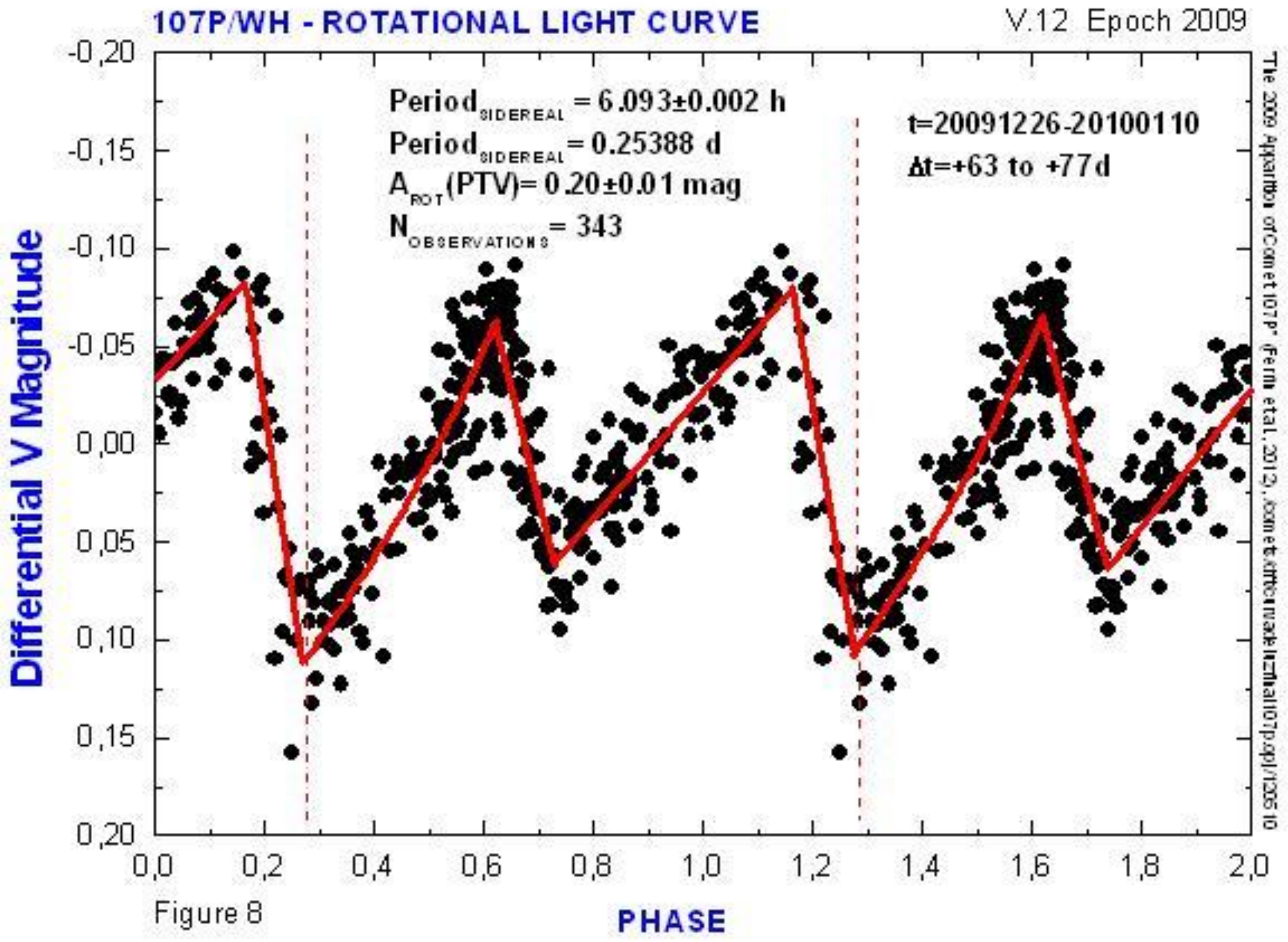

***Figure 8.*** *Rotational light curve of comet 107P/WH. The comet exhibits a well defined rotational modulation with a sidereal period of 6.093±0.002 h, and amplitude of 0.20±0.01 magnitudes. The ratio of axis a/b = 1.20±0.05. Notice the extreme slope at phase 0.2 which implies a change of 0.20 magnitudes in only 0.55 hours. The shape of the rotational light curve is that of a saw tooth. This implies that the shape of the nucleus must be very odd and uncommon and have sharp edges.*

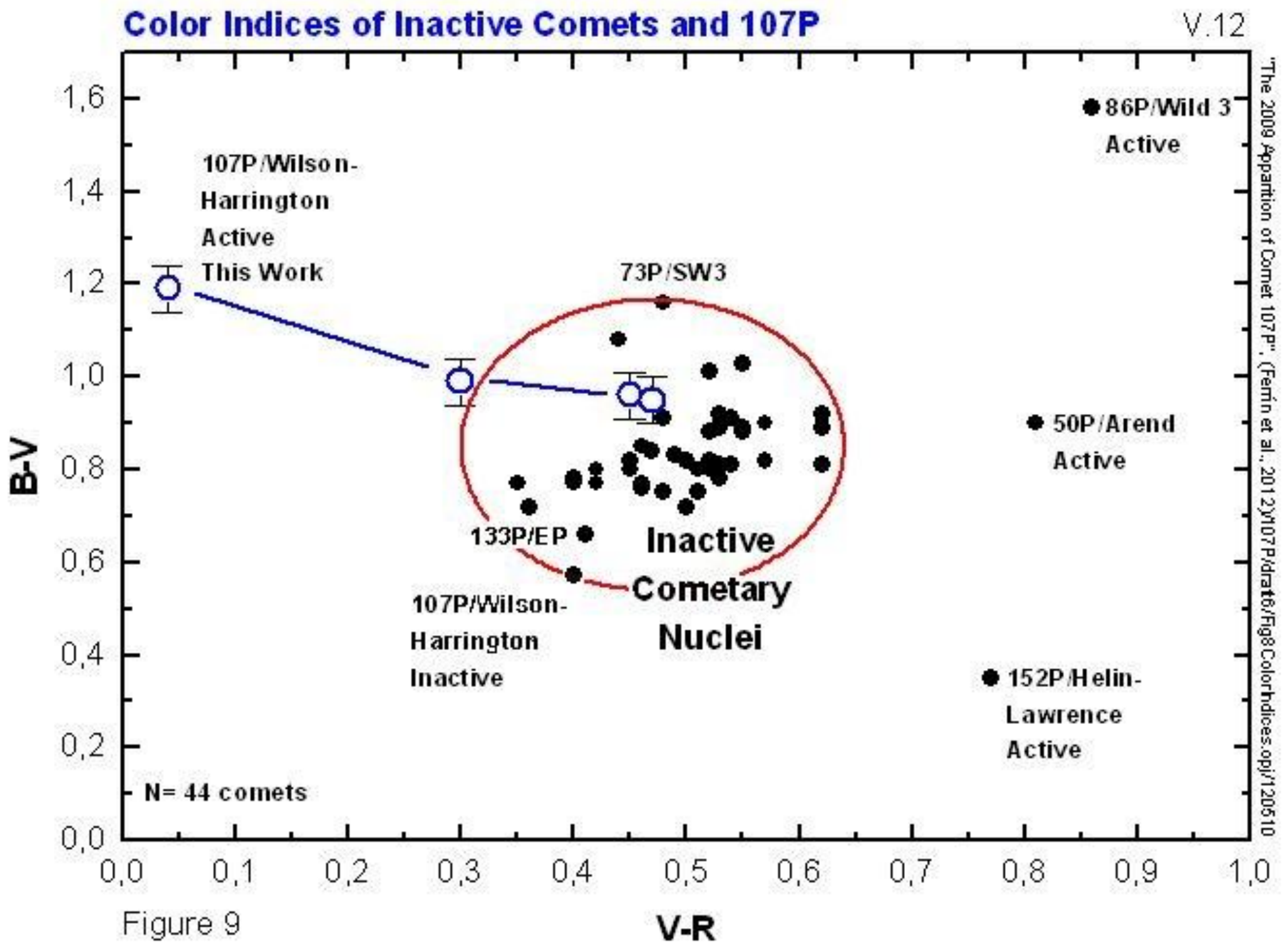

***Figure 9.*** *Color-Color diagram B-V vs V-R. The great majority of comets lie inside a circle, while 50P, 86P and 152P were active at the time of measurement. The data for this plot comes from Paper IV.*

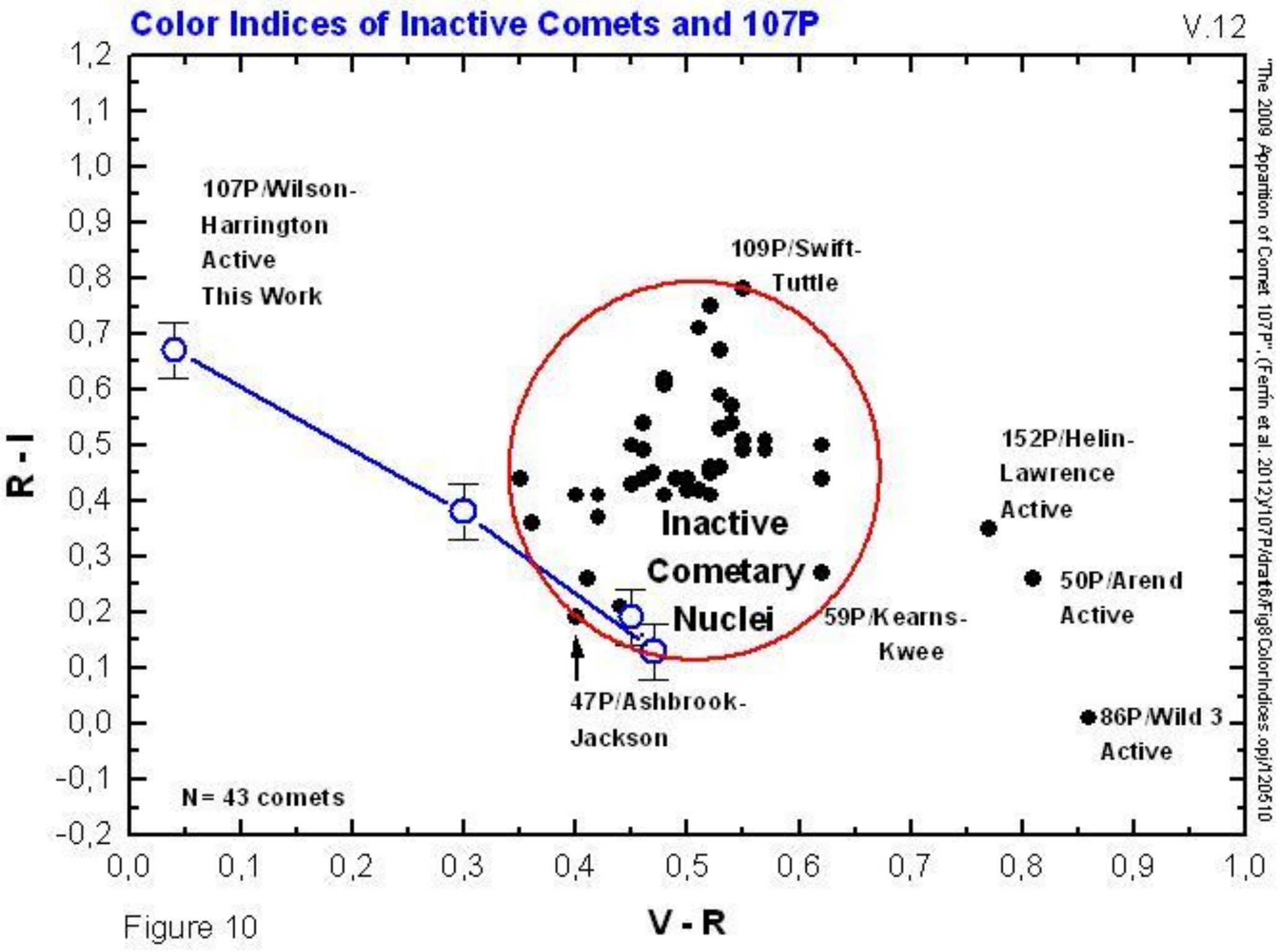

***Figure10.*** *Color-Color diagram R-I vs V-R. The great majority of comets lie inside a circle, while 50P, 86P and 152P were active at the time of measurement. The data for this plot comes from Paper IV.*

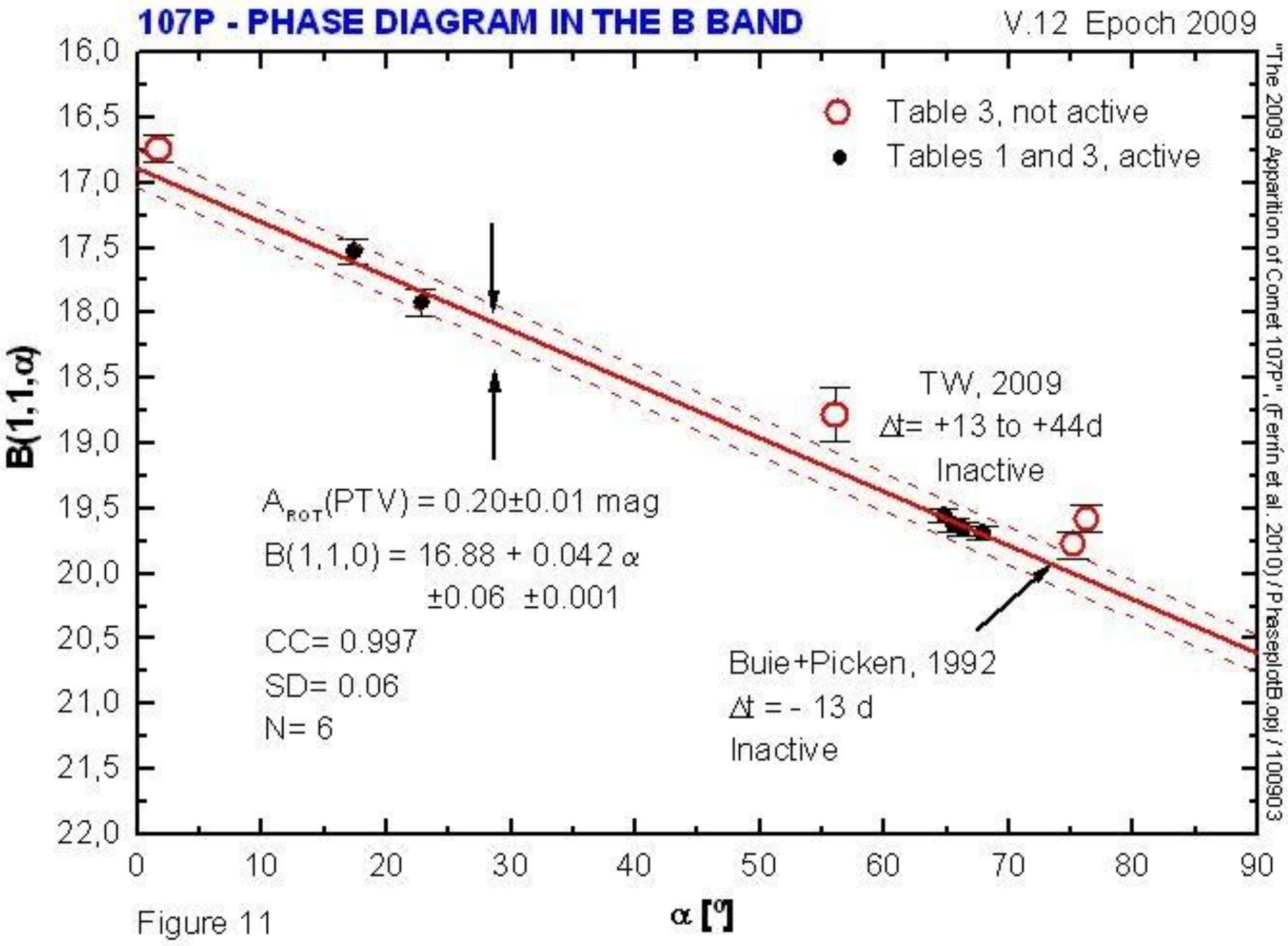

***Figure 11.*** *Phase plot of 107P in the blue B band. The data comes from Tables 1 and 3. The linear law is well defined. The comet was not active in the B band. TW = This Work.*

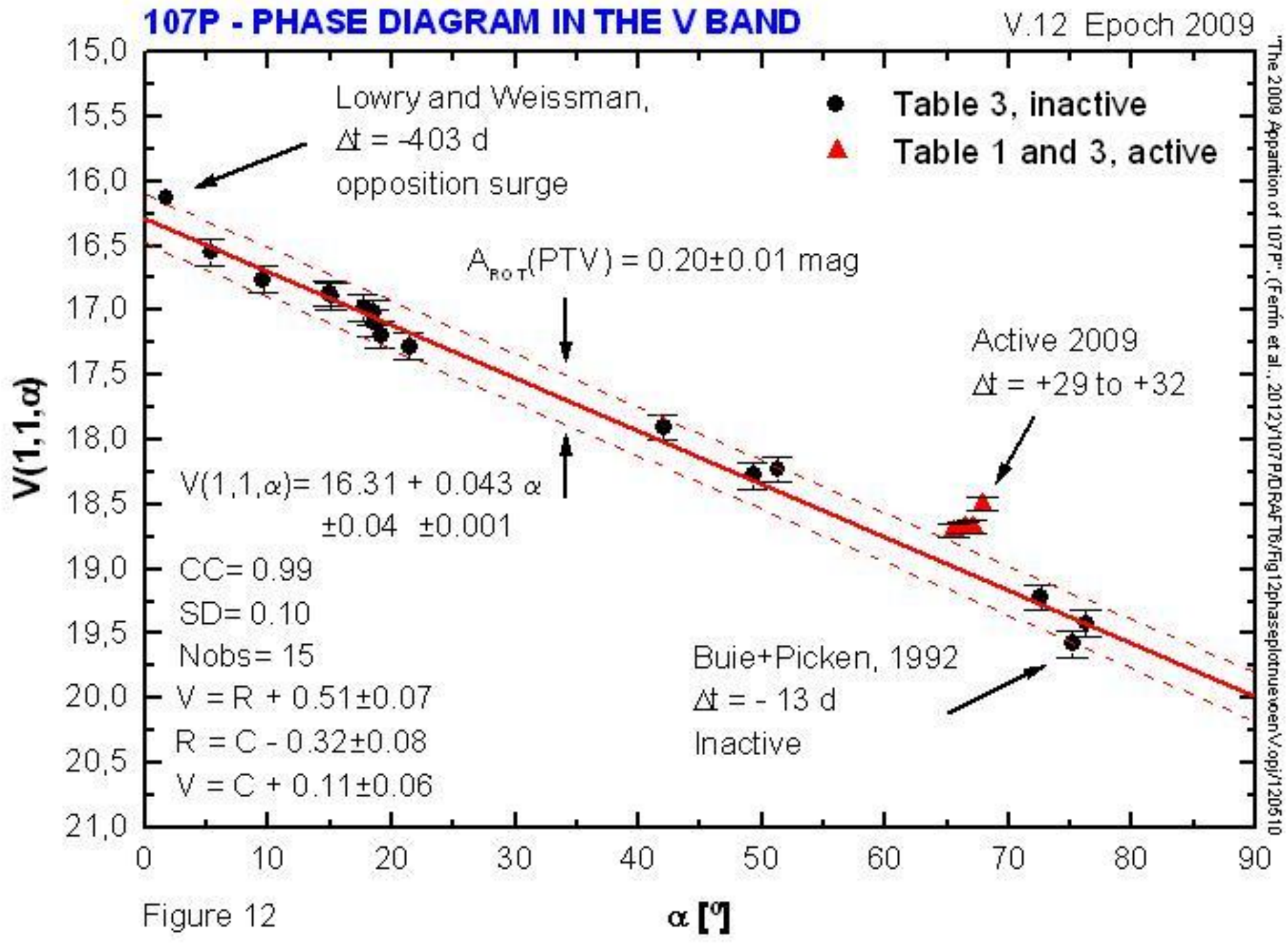

***Figure 12.*** *Phase plot of 107P in the visual V band. The data comes from Tables 1 and 3. The comet was not active in the V band although it may have been active marginally the last day of observations.*

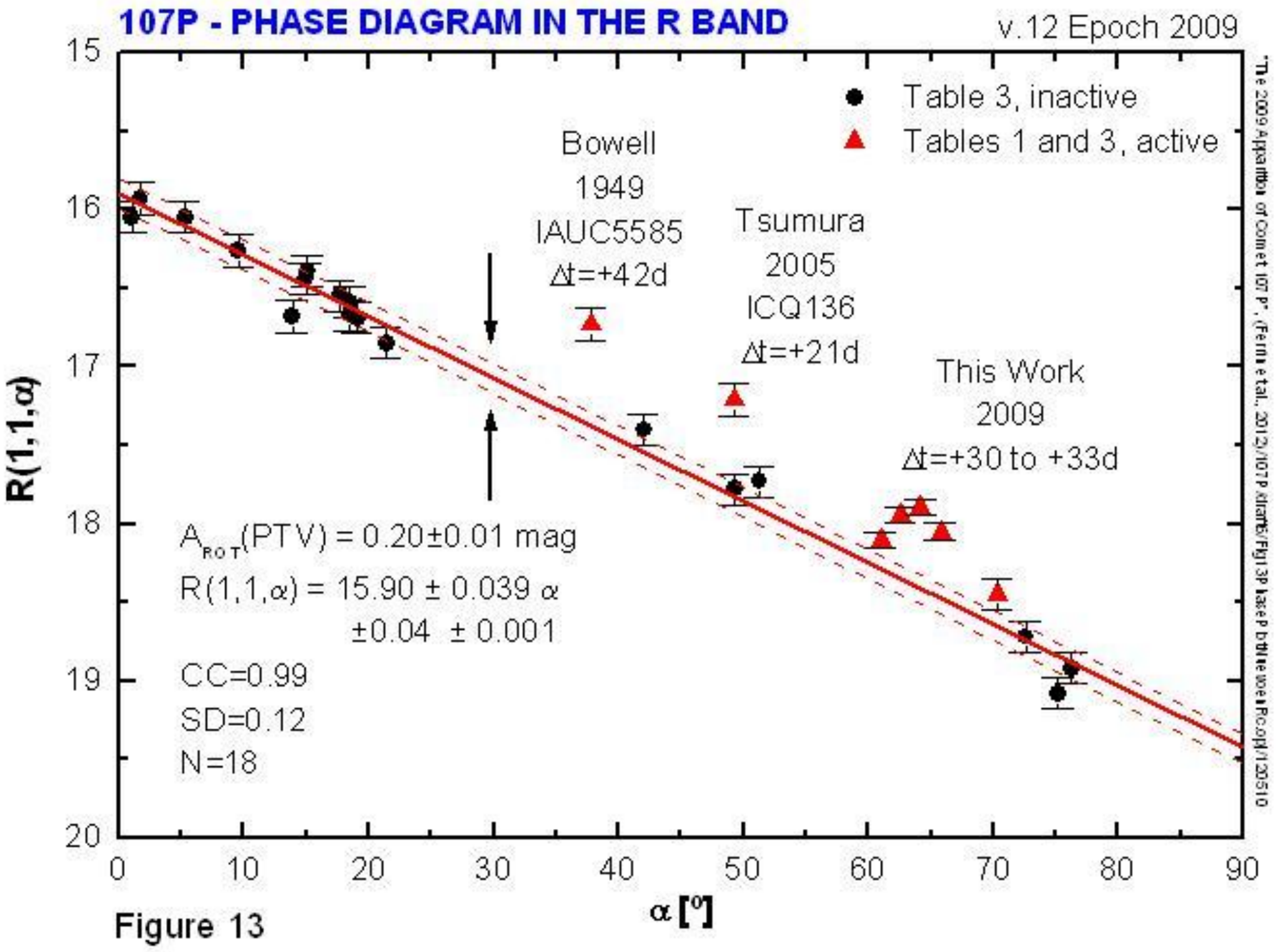

***Figure 13.*** *Phase plot of 107P in the red R band. The data comes from Tables 2 and 4. The phase line is well defined and our photometry was several σ above the nuclear line and beyond half the rotational amplitude. The comet was active in the R band in the five observing dates. TW = This Work.*

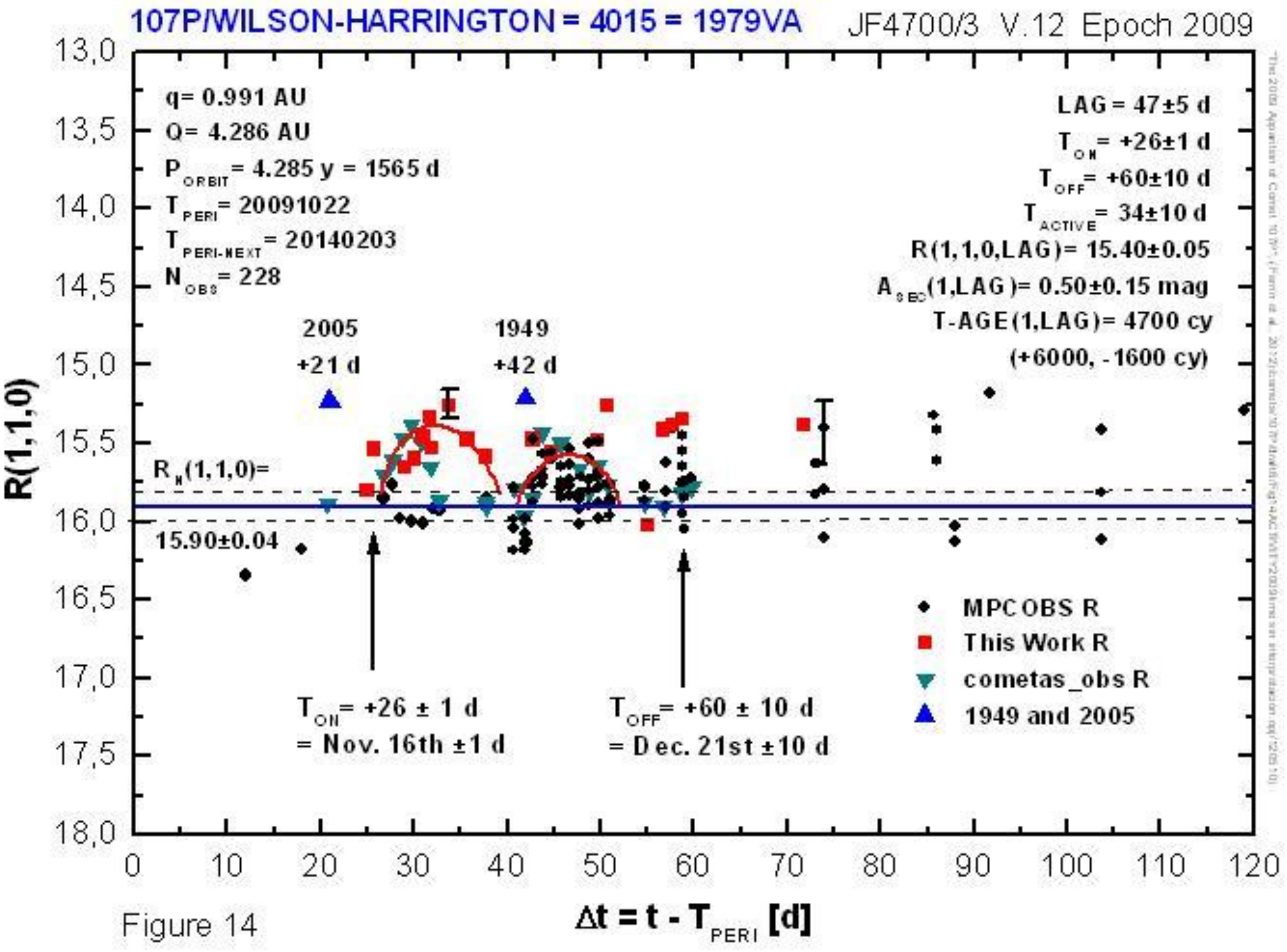

***Figure 14.*** *Secular light curve of 107P, time plot. The plot shows two active dates from 1949 and 2005 taken from Table 3. There is some scatter but the increase in magnitudes is more evident near the turn on time where two independent data set have detected an increase. Two active episodes have been suggested, but this may be an artifact due to uneven sampling.*

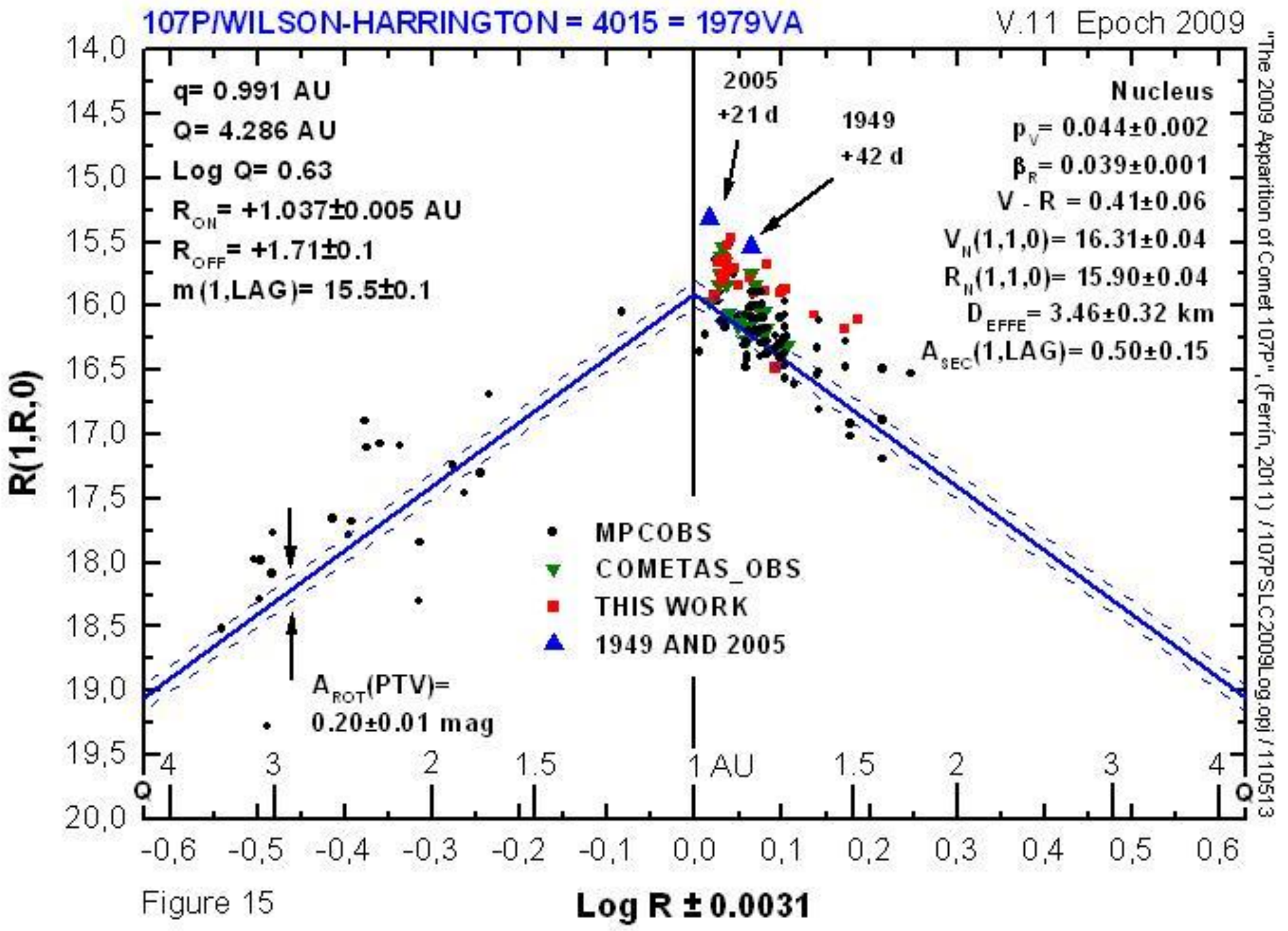

***Figure 15.*** *Secular light curve of 107P, log plot. Negative logs in the abscissa do not mean that the value is less than one, but that the distances are before perihelion. Notice that the plot covers the whole orbit from –Q to +Q. The comet was highly active in the two observations 1949 and 2005.*

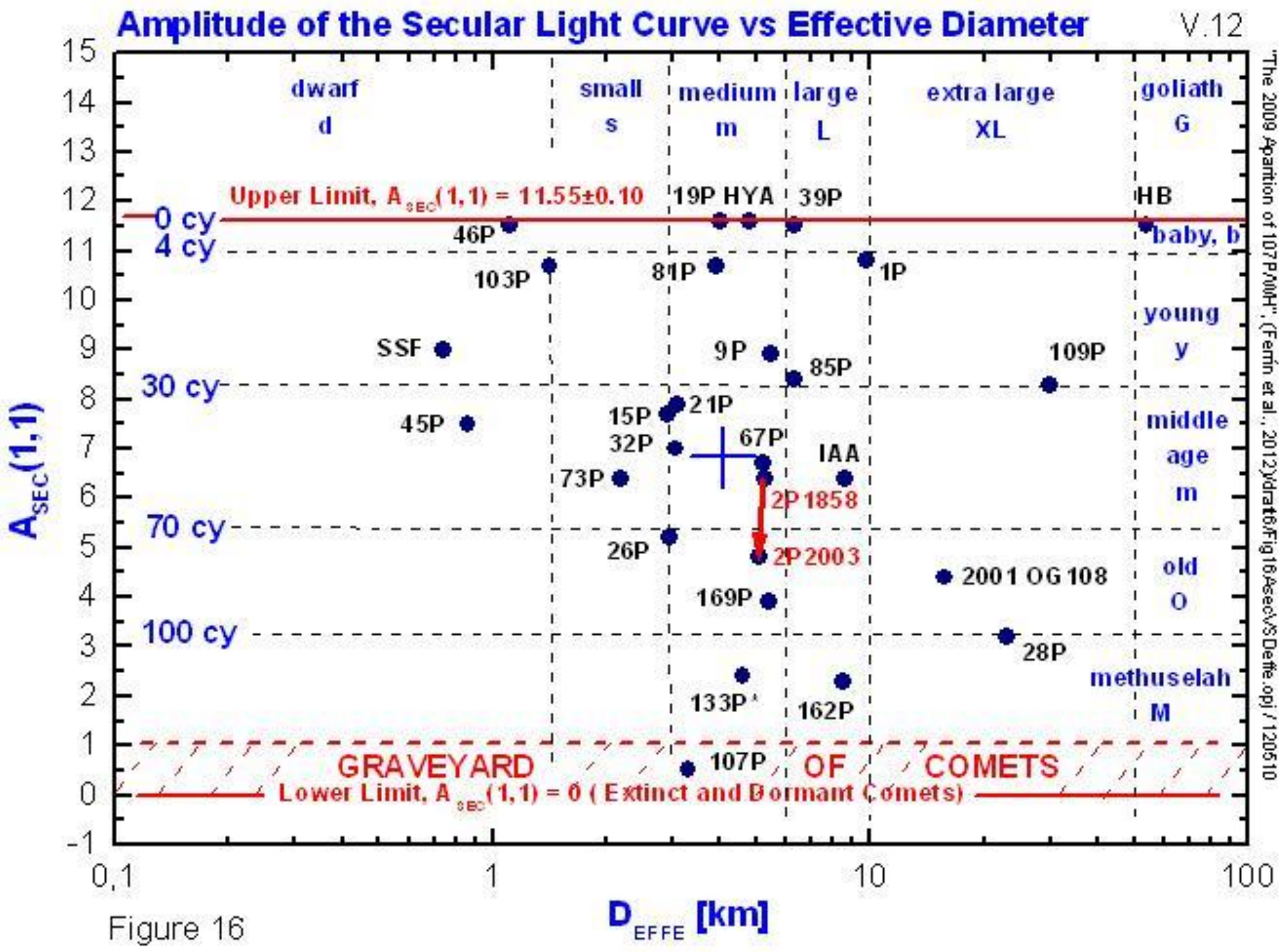

***Figure 16.*** *Amplitude of the secular light curve,* $A_{SEC}$*, vs effective diameter,* $D_{EFFE}$*, for 28 comets.* $A_{SEC}$ *is the height of the SLC at R=1 AU and it is not a function of diameter. Small comets are as active as large ones: Goliath comet Hale Bopp is as active as dwarf comet 46P/Wirtanen. Since comets decrease in activity as they age, and since they lose volatiles, comets move down and to the left in this diagram. In other words, this is an evolutionary diagram. Notice how comet 2P has changed position between the 1858 and 2003 (Paper VI). The location of 107P is at the bottom of the diagram. This implies that 107P is an extreme object, the most evolved of the whole sample.* $A_{SEC} = 0$ *implies an extinct or dormant comet. Since the general flow of the data set is down and to the left of this diagram, and since this flow has been going on for eons, and since dwarf comets evolve much more rapidly than large comets, this diagram predicts the existence of a significant enhanced population of dormant and extinct bodies located in the lower left region of the diagram. We call this region the graveyard of comets (* $0 < A_{SEC} < 1$ *mag). Comet 107P is the first member of this group.*

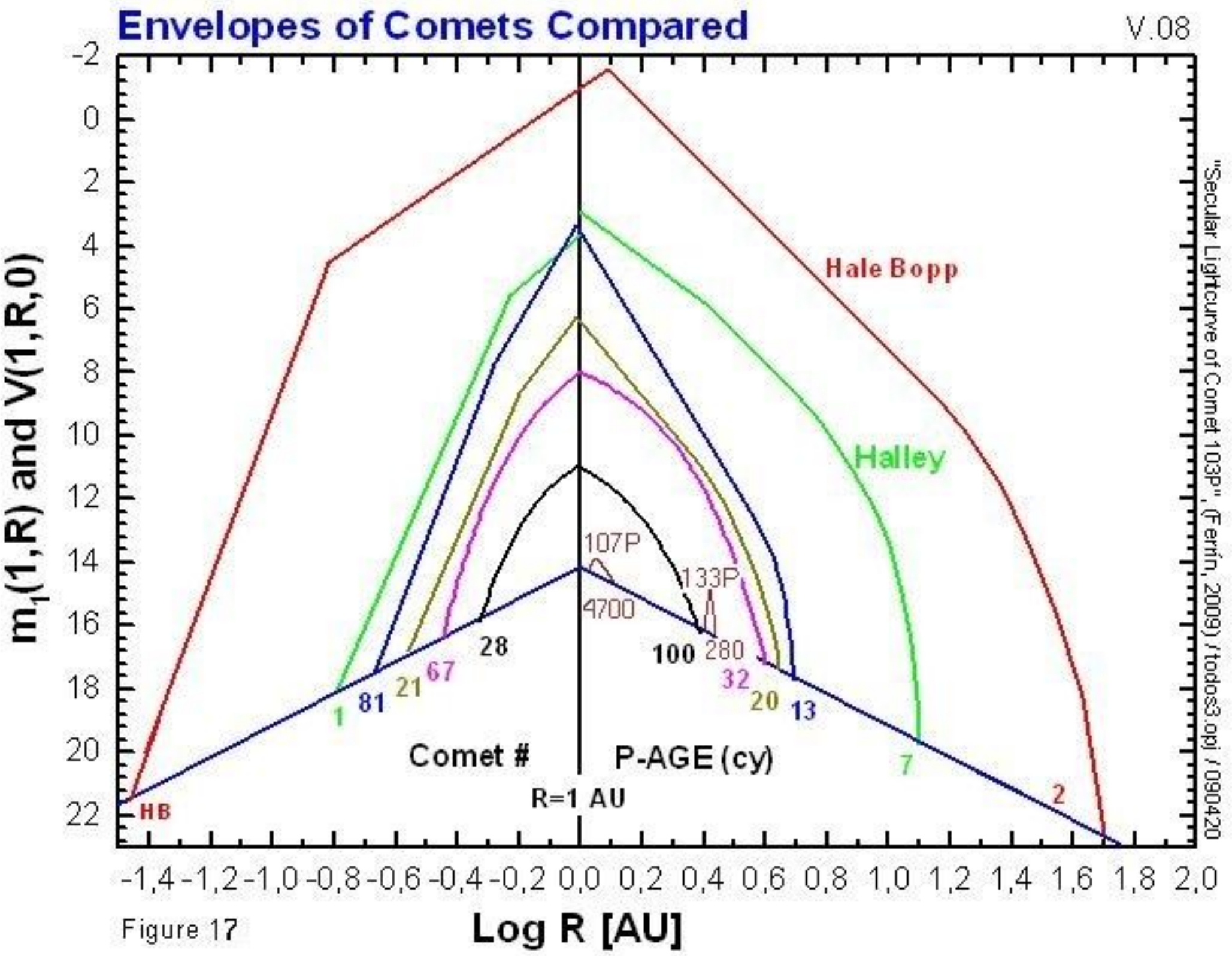

***Figure 17.*** *Validation of the concept of photometric age. Envelopes of the secular light curves of several comets are compared taken from Paper VII. Older comets are nested inside younger objects. In other words, P-AGE classifies comets by shape of the SLC. Additionally this figure shows that as a function of age, $A_{SEC}$ and $R_{SUM}$, diminish in value. A comet has to get nearer to the sun to get activated, and it is less and less active as it ages. P-AGE and T-AGE measure the activity of a comet and that activity diminishes monotonically as a function of age. The three oldest comets, 107P, 133P and D/1891 W1 Blanpain, turn on after perihelion. Thus there seems to be a jump from turn on before perihelion to turn on after perihelion for old comets.*

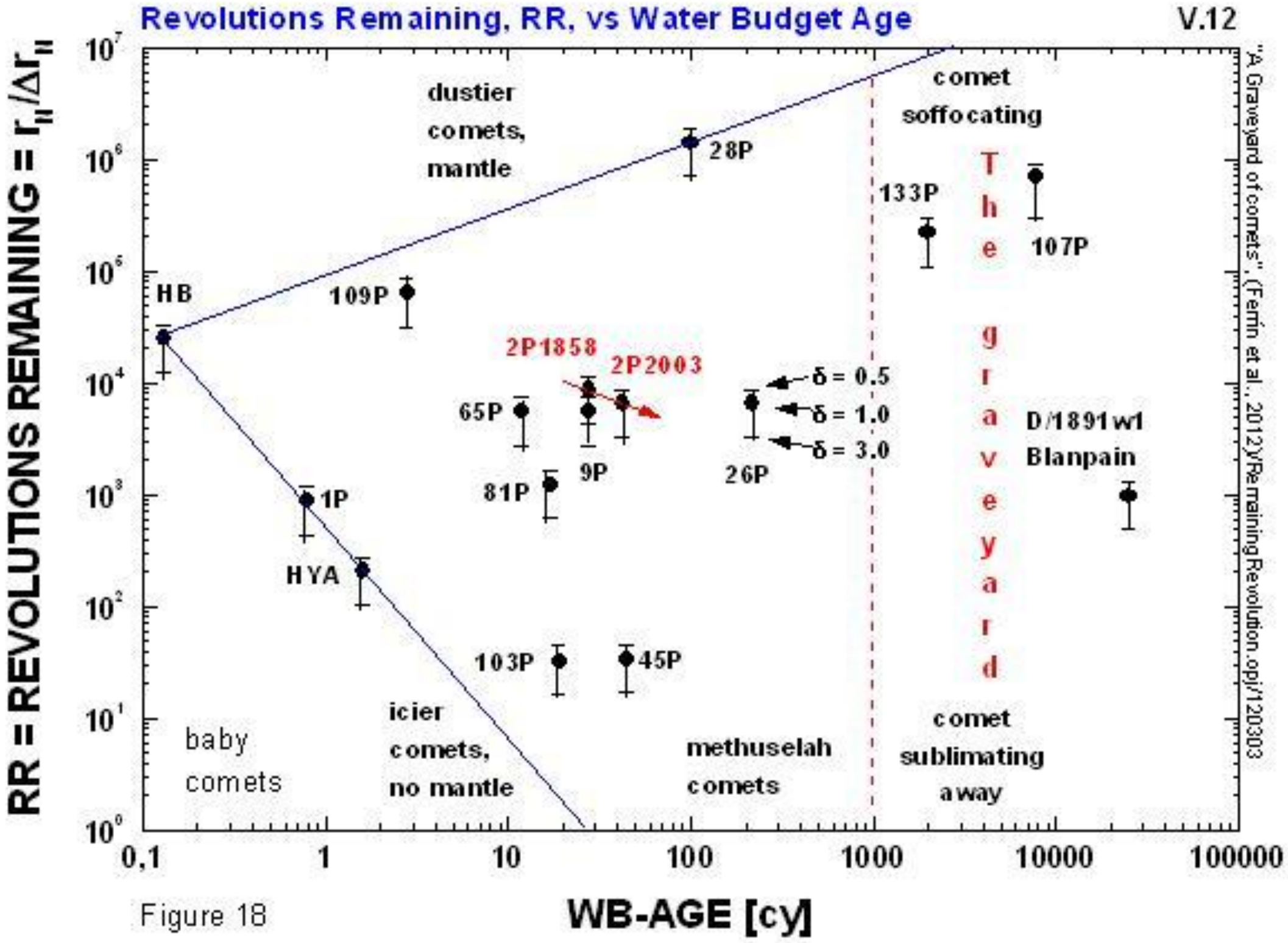

***Figure 18.*** *Remaining revolutions vs water budget age. This diagram is notorious in that the vertical and the horizontal axis each span a range of 7 and 6 orders of magnitude respectivelly. This is the range of activity of this sample of comets in the two parameter phase space. Comet 2P/Encke shows evolution of the parameters between the 1858 and the 2003 apparitions. Thus this is an evolutionary diagram. It is believed that comets evolve approximately from the location of comet Hale-Bopp, toward the right. If they move up they are choked by a dust crust. If they move down they sublimate away. 107P is the most extreme object in the upper right corner. The location of the comets is not sensitive to the dust to gas mass ratio. Two comets are sublimating away very rapidly, 103P and 45P. Three comets belong to the graveyard in this definition, 107P, 133P and D/1891W1 Blanpain.*